\documentclass[useAASTex,usenatbib]{mn2e}
\usepackage[english]{babel}
\usepackage{fleqn}
\usepackage{amsfonts}
\usepackage{amsmath}
\usepackage{amssymb}
\usepackage{hyperref}
\usepackage{booktabs}
\usepackage{pgfplots}
\usepackage{graphicx}
\usepackage{txfonts}

\begin{document}
\title[Triple population synthesis and white dwarf mergers]{Population synthesis of triple systems in the context of mergers of carbon-oxygen white dwarfs}
\author[A.S. Hamers, O.R. Pols, J.S.W. Claeys and G. Nelemans]{A.S. Hamers$^{1}$\thanks{E-mail: \tt{hamers@strw.leidenuniv.nl}}, O.R. Pols$^{2}$, J.S.W. Claeys$^{2}$ and G. Nelemans$^{2,3}$ \\
$^{1}$Leiden Observatory, Leiden University, PO Box 9513, NL-2300 RA Leiden, The Netherlands \\
$^{2}$Department of Astrophysics/IMAPP, Radboud University Nijmegen, PO Box 9010, NL-6500 GL Nijmegen, The Netherlands \\
$^{3}$Institute for Astronomy, KU Leuven, Celestijnenlaan 200D, 3001 Leuven, Belgium}

\date{Accepted for publication in MNRAS 2013 January 7.  Received 2013 January 7; in original form 2012 November 21}

\maketitle

\begin{abstract}
Hierarchical triple systems are common among field stars yet their long-term evolution is poorly understood theoretically. In such systems Kozai cycles can be induced in the inner binary system during which the inner orbit eccentricity and the inclination between both binary orbits vary periodically. These cycles, combined with tidal friction and gravitational wave emission, can significantly affect the inner binary evolution. To investigate these effects quantitatively we perform a population synthesis study of triple systems and focus on evolutionary paths that lead to mergers of carbon-oxygen (CO) white dwarfs (WDs), which constitute an important candidate progenitor channel for type Ia supernovae (SNe Ia). We approach this problem by Monte Carlo sampling from observation-based distributions of systems within the primary mass range $1.0 - 6.5 \, M_\odot$ and inner orbit semi-major axes $a_1$ and eccentricities $e_1$ satisfying $a_1 \left (1-e_1^2\right ) > 12 \, \mathrm{AU}$, i.e. non-interacting in the absence of a tertiary component. We evolve these systems by means of a newly developed algorithm that couples secular triple dynamics with an existing binary population synthesis code. We find that the tertiary significantly alters the inner binary evolution in about 24\% of all sampled systems. In particular, we find several channels leading to CO WD mergers. Amongst these is a novel channel in which a collision occurs in wide inner binary systems as a result of extremely high eccentricities induced by Kozai cycles. With assumptions on which CO WD mergers lead to a SN Ia explosion we estimate the SNe Ia delay time distribution resulting from triples and compare to a binary population synthesis study and to observations. Although we find that our triple rate is low, we have determined a lower limit of the triple-induced SNe Ia rate and further study is needed that includes triples with initially tighter inner orbits.
\end{abstract}


\section{Introduction}
\label{sect:introduction}
Coeval stellar hierarchical triple systems, henceforth triple systems, are known to be quite common among stellar systems in the field (e.g. \citealp{tok06, rag10}). Such systems consist of an inner binary pair orbited by a distant star, the tertiary. Due to secular gravitational effects high-amplitude eccentricity oscillations, known as Kozai or Kozai-Lidov cycles \citep{kozai62,lidov62}, can be induced in the inner binary system. These oscillations occur on timescales that are typically much longer than the orbital periods of both inner and outer binaries. One possible consequence of these eccentricity cycles is strong tidal friction and subsequent orbital shrinkage, a process which is known as Kozai cycles with tidal friction (KCTF). KCTF has been studied in the context of (solar-mass) main sequence (MS) stars \citep{maz79,egg01,kis10,fabrtr07}. In particular, \citet{fabrtr07} have shown that KCTF is responsible for producing close MS binary systems with periods $P_1 \sim 3 \, \mathrm{d}$, which is consistent with the observation that such systems are very likely ($96\%$) orbited by a tertiary \citep{tok06}. It remains to be seen, however, whether KCTF is still effective in higher-mass triples (individual masses $m \gtrsim 1.2 \, M_\odot$). This is because their constituents have radiative envelopes and are thus much less effective at dissipating tidal energy than their lower-mass counterparts, which have convective envelopes \citep{zahn77}. On the other hand, as such higher-mass stars greatly increase in size and develop convective envelopes during their red giant branch (RGB) and asymptotic giant branch (AGB) phases tidal friction is expected to become much more effective. Thus it is possible that KCTF is effective at significantly shrinking the inner binary system during the RGB/AGB phases, whereas this was not the case during the MS.

Another possible consequence of high-amplitude eccentricity cycles induced in (close) inner binary systems by a tertiary is the enhancement of the emission of gravitational waves (GWs), a process which is strongly dependent on the orbital eccentricity (e.g. \citealt{peters64}). \citet{thomp11} has suggested that this mechanism can significantly reduce the GW merger time in close carbon-oxygen (CO) white dwarf (WD) binaries if they are orbited by a tertiary with the appropriate orbital parameters. This scenario implies an increase in the expected number of close CO WD systems that merge within a Hubble time, which is relevant for binary population synthesis studies in which the CO WD merger channel is considered as an important candidate progenitor channel for type Ia supernovae (SNe Ia) (e.g. \citealt{ruiter09,toonen12}; Claeys et al. in prep.). 

In this work we investigate both scenarios of KCTF during stellar evolution and the merging of CO WDs under the influence of a tertiary companion by means of a population synthesis of triples. We take into account long-term dynamical, stellar and binary evolution by means of a newly developed algorithm in which a code modeling the secular dynamical evolution of triples is coupled with the binary population synthesis code {\tt Binary\_c} (see Sect. \ref{sect:triplealg:binalg} for details) to model the inner binary evolution. We restrict ourselves to intermediate mass triples with wide inner binary systems that would not interact in the absence of a tertiary component. For these triples we show in this paper that the presence of the tertiary component introduces to the inner binary system various novel evolutionary scenarios. In particular, we find an alternative pathway to SNe Ia as a result of collisions of WDs in wide inner binary systems in which the eccentricity reaches extremely high values as a result of Kozai cycles. 

The organization of this paper is as follows. In Sect. \ref{sect:triplealg} we describe the triple evolution algorithm developed in this work and in Sect. \ref{sect:extrev} we provide details of two example systems that lead to CO WD mergers via two distinct triple channels. With this algorithm we perform a population synthesis study of which we discuss our methods in Sect. \ref{sect:methods} and present the results in Sect. \ref{sect:results}. For the various CO WD merger channels we find we estimate the expected triple-induced SNe Ia rates and compare these to the binary population synthesis study of Claeys et al. (in prep.) and to rates inferred from observations in Sect. \ref{sect:implications}. In Sect. \ref{sect:discussion} we relate our results to the previous work of \citet{thomp11} and discuss the uncertainties of our treatment of triple evolution. We conclude in Sect. \ref{sect:conclusions}.

\section{Hierarchical triple evolution algorithm}
\label{sect:triplealg}

\subsection{Secular hierarchical triple dynamics}
\label{sect:triplealg:SHTD}
During Kozai cycles the mutual torque exerted by the inner and outer binary orbits causes an exchange of angular momentum between the orbits while the orbital energies, and thus the semi-major axes $a_j$, are conserved (e.g. \citealt{mard01}). Here we use index $j=1,2$ to denote quantities pertaining to the inner and outer orbits respectively. As a result of the exchange of angular momentum the orbital eccentricities $e_j$ and the total mutual inclination angle between both orbits $i_\mathrm{tot}$ vary periodically on timescales $P_K \gg P_2$, where $P_j$ denotes orbital period. To lowest order in $a_1/a_2$ $P_K$ is given by \citep{kinnak99}:
\begin{align}
P_K = \alpha \frac{P_2^2}{P_1} \frac{m_1+m_2+m_3}{m_3} \left(1-e_2^2\right)^{3/2},
\label{eq:PK}
\end{align}
where $m_1$ and $m_2$ denote the inner binary primary and secondary mass respectively and $m_3$ denotes the tertiary mass. The dimensionless quantity $\alpha$ is of order unity and depends weakly on the initial values of $i_\mathrm{tot}$, $e_1$ and $g_1$, where $g_j$ denotes the argument of periastron. Other quantities that vary periodically are $g_j$ (i.e. orbital precession, also known as apsidal motion) and the longitudes of the ascending nodes $h_j$. 

In systems with large $a_1/a_2$ the eccentric Kozai mechanism may apply in some cases (\citealt{lithnaoz11}, \citealt{shapthomp12}). This mechanism is associated with orbital ``flips" (change of orbital inclination from retrograde to prograde and vice versa) with potentially very high inner orbit eccentricities. A quantity that measures the importance of this effect is the ``octupole parameter'' $\epsilon_\mathrm{oct}$ defined by:
\begin{align}
\epsilon_\mathrm{oct} = \frac{m_1-m_2}{m_1+m_2} \frac{a_1}{a_2} \frac{e_2}{1-e_2^2},
\label{eq:eoct}
\end{align}
where typically $\log_{10} (|\epsilon_\mathrm{oct}|) \gtrsim -2$ indicates that the eccentric Kozai mechanism could be important \citep{naoz11,shapthomp12}. 
 
In physically realistic situations there exist sources of precession in the inner orbit other than those due to secular three-body dynamics alone (in our study such effects in the outer orbit are negligible because $a_2$ is always sufficiently large). We take into account the following additional contributions to precession: due to general relativistic effects ($\dot{g}_{1,\mathrm{GR}}$), due to tidal bulges induced in physically extended stars in relative close proximity ($\dot{g}_{1,\mathrm{tide}}$) and due to intrinsic spin rotation ($\dot{g}_{1,\mathrm{rotate}}$). The fact that $\dot{g}_{1, \mathrm{GR}}$, $\dot{g}_{1, \mathrm{tide}}$ and $\dot{g}_{1,\mathrm{rotate}}$ are all positive implies that the associated sources of precession promote precession in one particular direction and therefore suppress the resonance between $g_1$ and $e_1$ during Kozai cycles. This phenomenon has been investigated in detail by \citet{blaes02}. 

Furthermore, whenever the distances between the components in triple systems become comparable to their radii tidal effects may become important. In our study such effects in the outer orbit can be neglected. However, in the inner binary system tidal effects may become important even for large semi-major axes with respect to the radii, $a_1/R_i$ (where $R_i$ denotes the inner binary stellar radius), if the inner orbital eccentricity is sufficiently high. In this case a tidal torque is induced, either due to the equilibrium tide or dynamical tide \citep{zahn77}, that allows for the exchange between angular momenta of the spin and orbit in the inner binary system. This process continues until coplanarity, synchronization and circularization are achieved \citep{hut80}. It is described quantitatively in terms of orbit-averaged equations for the orbital parameter time derivatives, $\dot{a}_{1,\mathrm{TF}}$, $\dot{e}_{1,\mathrm{TF}}$ and $\dot{\Omega}_{i,\mathrm{TF}}$, where $\Omega_i$ is the spin frequency of inner orbit star $i$ and TF denotes tidal friction. 

Another effect we take into account is GW emission, which we model for the inner binary only. As a consequence of the emission of GWs the inner binary is circularized and shrinks, which is described quantitatively by orbit-averaged equations for $\dot{e}_{1,\mathrm{GW}}$ and $\dot{a}_{1,\mathrm{GW}}$. 

\subsection{System of first-order differential equations}
\label{sect:triplealg:ODEs}
We describe the secular Kozai cycles quantitatively in terms of coupled equations for $\dot{g}_{j,\mathrm{STD}}$ and $\dot{e}_{j,\mathrm{STD}}$ (where STD denotes secular three-body dynamics), that follow after orbital averaging of the three-body Hamiltonian expressed in Delaunay's action-angle variables, applying Hamilton's equations of motion and eliminating the longitudes of the ascending nodes \citep{har68,ford00,naoz11}. We employ equations for these four quantities accurate to octupole order, i.e. up to and including third order in $a_1/a_2$. In addition, an equation that expresses conservation of total angular momentum describes the change of $i_\mathrm{tot}$. 

We then combine all the physical effects discussed in Sect. \ref{sect:triplealg:SHTD} to form a system of eight first-order ordinary differential equations (ODEs) that reads:
\begin{align}
\left \{ \begin{array}{cl}
\dot{g}_1 &= \dot{g}_{1,\mathrm{STD}} + \dot{g}_{1,\mathrm{GR}} + \dot{g}_{1,\mathrm{tide}} + \dot{g}_{1,\mathrm{rotate}}; \\
\dot{g}_2 &= \dot{g}_{2,\mathrm{STD}}; \\
\dot{e}_1 &= \dot{e}_{1,\mathrm{STD}} + \dot{e}_{1,\mathrm{TF}} + \dot{e}_{1,\mathrm{GW}}; \\
\dot{e}_2 &= \dot{e}_{2,\mathrm{STD}}; \\
\dot{a}_1 &= \dot{a}_{1,\mathrm{TF}} + \dot{a}_{1,\mathrm{GW}}; \\
\dot{\theta} &= \frac{-1}{G_1G_2} \left [ \dot{G}_1 \left (G_1 + G_2 \theta \right ) + \dot{G}_2 \left (G_2 + G_1 \theta \right ) \right ]; \\
\dot{\Omega}_1 &= \dot{\Omega}_{1,\mathrm{TF}}; \\
\dot{\Omega}_2 &= \dot{\Omega}_{2,\mathrm{TF}},
\end{array} \right.
\label{eq:tripleODEs}
\end{align}
where $\theta \equiv \cos(i_\mathrm{tot})$ and $G_j$ denotes the orbital angular momentum. For expressions for the precession quantities $\dot{g}_{1,\mathrm{GR}}$, $ \dot{g}_{1,\mathrm{tide}}$ and $\dot{g}_{1,\mathrm{rotate}}$ we refer to \citet{blaes02}, \citet{smwil01} and \citet{fabrtr07} respectively; for the tidal evolution quantities $\dot{e}_{1,\mathrm{TF}}$, $\dot{a}_{1,\mathrm{TF}}$ and $\dot{\Omega}_{i,\mathrm{TF}}$ we refer to \citet{hut81} and for the GW evolution quantities $\dot{e}_{1,\mathrm{GW}}$ and $\dot{a}_{1,\mathrm{GW}}$ we refer to \citet{peters64}. The main assumption in Eq. \ref{eq:tripleODEs} is that all physical processes that are described are independent such that the individual time derivative terms can be added linearly. In addition, in the expressions for $\dot{g}_{1,\mathrm{tide}}$, $\dot{g}_{1,\mathrm{rotate}}$, $\dot{e}_{1,\mathrm{TF}}$, $\dot{a}_{1,\mathrm{TF}}$, $\dot{\Omega}_{i,\mathrm{TF}}$ and $\dot{\theta}$ we assume coplanarity of spin and orbit at all times, even though Kozai cycles in principle affect the relative orientations between the spin and orbit angular momentum vectors and in turn a misalignment of these vectors affects the Kozai cycles themselves (e.g. \citealt{correia10}). We justify this assumption by noting that for the majority of systems that we study the orbital angular momenta of both inner and outer orbits greatly exceed the spin angular momenta in magnitude, therefore the stellar spins cannot greatly affect the exchange of angular momentum between both orbits. 

The expression for $\dot{\theta}$ in Eq. \ref{eq:tripleODEs} is derived by differentiating the relation that expresses conservation of total angular momentum, $\boldsymbol{G}_\mathrm{tot}^2 = G_1^2 + G_2^2 + 2 \, G_1 G_2 \theta$, with respect to time and solving for $\dot{\theta}$ in terms of $G_j$ and $\dot{G}_j$. Note that GW emission acts to change $a_1$ and $e_1$ and therefore affects the magnitude of $\boldsymbol{G}_1$ but not its direction. Although in principle tidal friction can also affect the direction of $\boldsymbol{G}_1$ if the stellar spins and orbit are not coplanar, with our assumption of coplanarity this is not the case. This implies that, with $G_j = L_j \sqrt{1-e_j^2}$, where $L_j = L_j(a_j,m_i)$ is the orbital angular momentum in case of a circular orbit and which is constant with regard to STD, the quantities $\dot{G}_j$ are given by $\dot{G}_j = - \left [ e_j/\left(1-e_j^2\right) \right ] \cdot G_j \cdot \dot{e}_{j,\mathrm{STD}}$. 

\subsection{Coupling to binary algorithm}
\label{sect:triplealg:binalg}
The hierarchical triple evolution algorithm developed in this work couples the system of differential equations Eq. \ref{eq:tripleODEs} to the existing rapid binary population synthesis algorithm {\tt Binary\_c} based on \citet{hur00,hur02} with updates described in \citet{izz04, izz06,izz09} and Claeys et al. (in prep.). This binary star algorithm includes stellar evolution and a variety of binary interaction processes such as mass loss due to stellar winds, mass transfer and common envelope (CE) evolution. We refer to \citet{hur02} for further details of this binary population synthesis algorithm. The novel aspect of our approach is that the evolution of the inner binary system is taken into account through the coupling to {\tt Binary\_c}. We also take into account the evolution of the tertiary, which is treated as being isolated such that its mass may be computed from single stellar evolution. In this work we consider systems in which at all times the tertiary is sufficiently distant from the inner binary system such that it does not fill its Roche-lobe in its orbit around the inner binary system, nor it is involved in a CE phase with the inner binary system, a scenario which has been explored in earlier studies (e.g. \citealt{ibtut99}). Therefore the latter processes are not modeled in the triple algorithm.

In the ODE solver routine the system of differential equations Eq. \ref{eq:tripleODEs} is solved for the duration of each time step of the {\tt Binary\_c} algorithm. In order to ensure proper convergence of the solutions of Eq. \ref{eq:tripleODEs}, which is quite stiff in nature, we use an ODE solver specifically designed to integrate stiff ODEs \citep{co96}. We have checked the results of the ODE solver routine when not coupled with the {\tt Binary\_c} algorithm by computing the evolution of the same hierarchical triple systems as studied in \citet{ford00}, \citet{blaes02} and \citet{naoz11} and find very similar results for $e_1$, $i_\mathrm{tot}$ and $a_1$ (where applicable) as function of time. We have also found near exact agreement of the time evolution of $e_1$ and $i_\mathrm{tot}$ when comparing to analytical solutions which exist in the quadrupole order approximation and with the further restriction that $G_2 \gg G_1$ \citep{kinnak99}. During each {\tt Binary\_c} time step quantities not included in the left hand side of Eq. \ref{eq:tripleODEs} such as the inner binary masses and radii are assumed to be constant. Because the {\tt Binary\_c} algorithm uses adaptive time steps to match the rate of stellar and binary evolution it is always ensured that these quantities do not change significantly during each iteration of the triple evolution algorithm. In addition, if needed the triple algorithm decreases the {\tt Binary\_c} time step such that the inner orbital angular momentum $G_1$ does not change by more than $2 \%$. Thus, changes to the inner orbit semi-major axis and eccentricity due to secular three-body dynamics coupled with tidal friction and GW emission are communicated to the {\tt Binary\_c} algorithm at appropriate times. 

The version of {\tt Binary\_c} used in this work enforces circularization at the onset of Roche-lobe overflow (RLOF) and CE evolution. Although this assumption is generally justified for isolated binaries \citep{hur02}, in triple systems high-amplitude eccentricity cycles could be induced during the former phase, in particular if the timescale of these cycles is comparable to or smaller than the timescale of mass transfer driven by RLOF. An accurate treatment of mass transfer in eccentric orbits is beyond the scope of this work (see e.g. \citealt{sep07}) and we therefore choose to disable the ODE solver routine whenever mass transfer driven by RLOF or CE evolution ensues. Whenever the ODE solver routine is active it is ensured that tidal evolution and GW emission in the inner binary system as calculated by {\tt Binary\_c} do not change the inner orbital parameters. This is necessary because in this case these processes are taken into account by the ODE solver routine through the equations for $\dot{e}_1$ and $\dot{a}_1$ in Eq. \ref{eq:tripleODEs}. In this manner the coupling of eccentricity cycles with tidal friction and GW emission in the inner orbit is calculated consistently.

If the ratio $\beta \equiv a_2/a_1$ is large the effects of STD are likely to be quenched by different processes in the inner binary system, in particular due to general relativistic precession in very tight inner binary systems, as discussed in Sect. \ref{sect:triplealg:SHTD}. In such situations it is not necessary to solve Eq. \ref{eq:tripleODEs} because the STD-related terms are negligible and the inner binary evolution can therefore be fully handled by {\tt Binary\_c}. We therefore introduce a criterion that temporarily disables the ODE solver routine and leaves the inner binary evolution to {\tt Binary\_c} in such cases. Specifically, the routine is disabled if $\beta > 10 \, \beta_\mathrm{crit,GR}$, where $\beta_\mathrm{crit,GR}$ is the value of $\beta$ such that the periods associated with precession due to STD (Eq. \ref{eq:PK}) and general relativity (approximately given by the reciprocal of $\dot{g}_{1,\mathrm{GR}}$) are equal, which we find is given by (omitting numerical factors of order unity): 
\begin{align}
\nonumber \beta_\mathrm{crit,GR} &\approx \left ( \frac{c^2}{G_N} \right )^{1/3} a_1^{1/3} \left( \frac{m_3}{(m_1+m_2)^2} \right )^{1/3} \frac{\left(1-e_1^2\right)^{1/3}}{\left(1-e_2^2\right)^{1/2}} \\ &\approx 5 \cdot 10^2 \left ( \frac{a_1}{\mathrm{AU}} \right )^{1/3} \left ( \frac{m_3/M_\odot}{\left(m_1/M_\odot + m_2/M_\odot \right )^2} \right )^{1/3}  \frac{\left(1-e_1^2\right)^{1/3}}{\left(1-e_2^2\right)^{1/2}},
\label{eq:betacritGR}
\end{align}
where $c$ denotes the speed of light and $G_N$ denotes the gravitational constant.

Outer orbital expansion due to mass loss in either inner or outer binary systems is taken into account with the assumption of fast, isotropic winds (i.e. Jeans mode) such that $a_2(m_1+m_2+m_3)$ and $e_2$ remain constant \citep{huang56,huang63}. With this relation a new value of $a_2$ is computed from the changes of $m_1$, $m_2$ and $m_3$ during each time step of {\tt Binary\_c}. 

Lastly, at each {\tt Binary\_c} time step the triple system is checked for dynamical stability by means of the stability criterion formulated by \citet{mard01}, including the ad hoc inclination factor $f = 1- (0.3/\pi) \, i_\mathrm{tot}$ (with $i_\mathrm{tot}$ expressed in radians). Whenever $\beta \leq \beta_\mathrm{crit}$, where $\beta_\mathrm{crit}$ is given by this stability criterion, the STD equations are no longer strictly valid we do not model the subsequent evolution. The phase of dynamical instability would need to be computed with a three-body integrator code, which is beyond the scope of this work (see \citealt{perkr12} for a recent study of triple systems that become dynamically unstable).

\begin{figure*}
\center
\includegraphics[scale = 1.4, trim = 0mm 1mm 0mm 0mm]{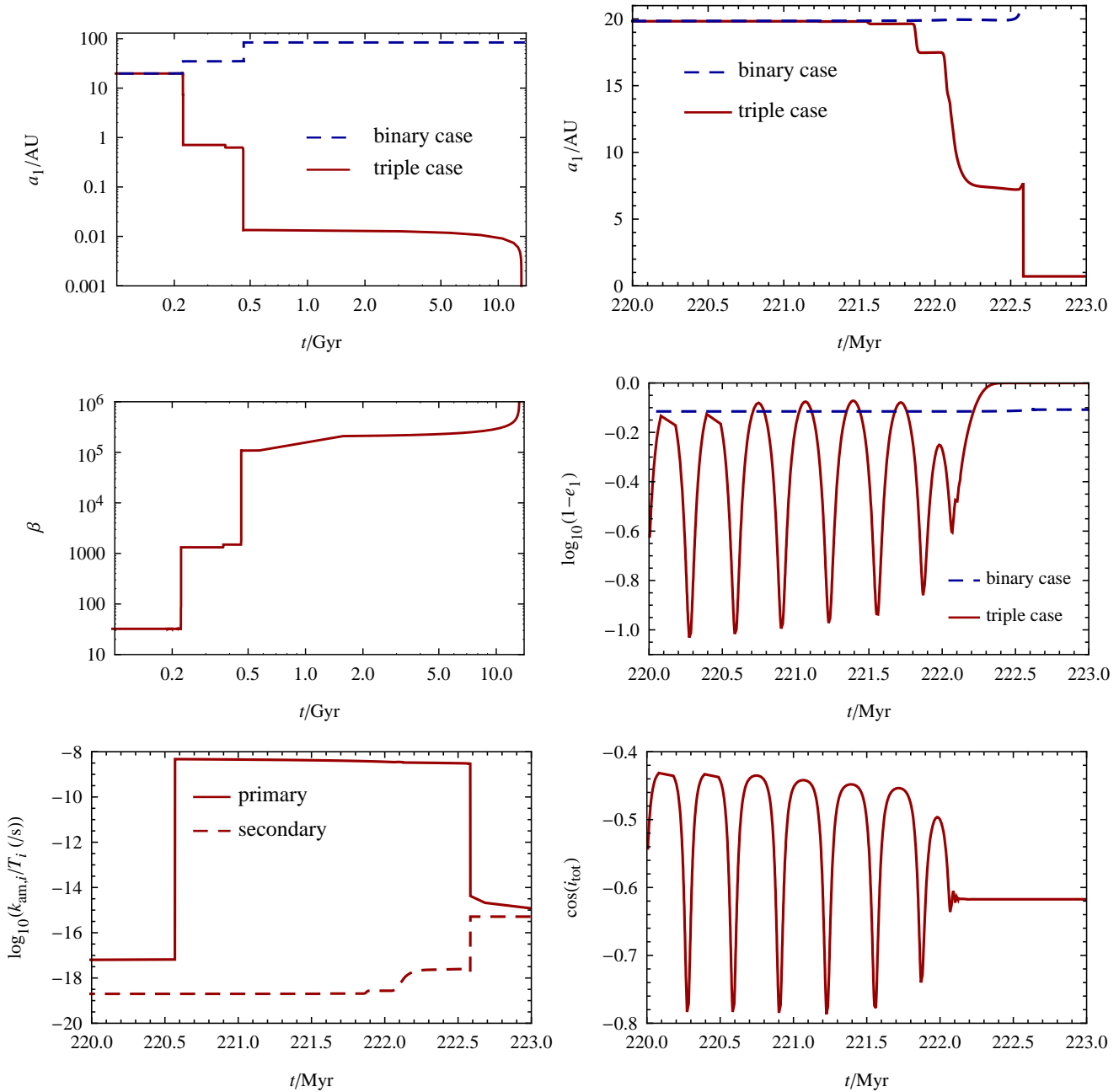}
\caption{\small Several quantities of interest in the evolution of the first example system (Sect. \ref{sect:extrev:TICM}). Shown as a function of time are $a_1$, $e_1$, $\beta \equiv a_2/a_1$, $k_{\mathrm{am},}i/T_i$ and $i_\mathrm{tot}$. In the plots in the first column of the first two rows the entire evolution is shown; the other plots correspond to the end of the primary core helium burning phase, the primary AGB phase and start of the primary CO WD phase. Solid line: triple case; dashed line: binary case (where applicable). In the plot of $k_{\mathrm{am},}i/T_i$ only the triple case is shown (the binary case is very similar); in this plot the solid line applies to the primary ($i=1$) and the dashed line to the secondary ($i=2$). Note that the evolution is not fully sampled in all plots, causing several kinks (in particular in the plots for $e_1$ and $i_\mathrm{tot}$) -- more detailed calculations are performed internally in the ODE solver routine but are not shown here (cf. Sect. \ref{sect:triplealg:binalg}).}
\label{fig:ex10131}
\end{figure*}

\subsection{Fixed binary parameters and evaluation of various physical quantities}
\label{sect:triplealg:physquan}
We detail some of the fixed binary evolution parameters used in this work. We choose quasi-solar metallicity, i.e. $Z = 0.02$. The initial spin periods of the inner orbit components are computed from a formula given by \citet{hur00}, which is fitted from data of the rotational speed of MS stars of \citet{lang92}. The CE parameter $\alpha_\mathrm{CE}$ describes the efficiency of the conversion of orbital energy into binding energy with which to shed the envelope and it is set to the canonical value of $\alpha_\mathrm{CE} = 1$. In this CE description an additional parameter $\lambda_\mathrm{CE}$ is required, which is a dimensionless measure of the relative density distribution within the envelope; here it is determined by means of functional fits as detailed in Claeys et al. (in prep.). All other parameters intrinsic to the {\tt Binary\_c} algorithm are also set to identical values as in Claeys et al. (in prep.).

In addition, various physical quantities are required to evaluate the right-hand sides in Eq. \ref{eq:tripleODEs} and in the remainder of this section we describe how these quantities are obtained in our triple algorithm. The tidal evolution equations (i.e. expressions for $\dot{a}_{1,\mathrm{TF}}$, $\dot{e}_{1,\mathrm{TF}}$ and $\dot{\Omega}_{i,\mathrm{TF}}$) contain a parameter that captures the efficiency of tidal friction and depends on the structure of the star \citep{hut81}. This parameter is the ratio $k_{\mathrm{am},i}/T_i$, where $k_{\mathrm{am},i}$ is the classical apsidal motion constant of inner orbit star $i$ and $T_i$ a tidal friction timescale. We compute $k_{\mathrm{am},i}/T_i$ according to the same prescription that is used in {\tt Binary\_c} \citep{hur02}. In this prescription a distinction is made between convective, radiative and degenerate envelopes and $k_{\mathrm{am},i}/T_i$ for these three cases is computed based on results of \citet{rasio96}, \citet{zahn77} and \citet{camp84} respectively. Furthermore, the gyration radii $r_{g,i}^2$ are required for the evaluation of $\dot{\Omega}_{i,\mathrm{TF}}$ and are computed with a prescription in which the stars are split into two parts, consisting of the core and the envelope, detailed in \citet{hur00} and \citet{hur02}. The quantities $k_{\mathrm{am},i}/T_i$ and $r_{g,i}^2$ are therefore calculated from precisely the same prescription as in {\tt Binary\_c}, which is required for a consistent treatment. 

The classical apsidal motion constant $k_{\mathrm{am},i}$ (i.e. not the {\it ratio} $k_{\mathrm{am},i}/T_i$), which is required for evaluation of $\dot{g}_{1,\mathrm{tide}}$ and $\dot{g}_{1,\mathrm{rotate}}$ \citep{smwil01,fabrtr07}, is computed as a function of mass and time relative to the ZAMS from detailed stellar models for metallicity $Z=0.02$ calculated by \citet{claret04}. The run of $k_{\mathrm{am},i}$ with time is determined by means of linear relations scaled to the time spent during each evolutionary stage for all tabulated masses in \citet{claret04}. Subsequently, $k_{\mathrm{am},i}$ is calculated for arbitrary mass by means of linear interpolation between adjacent mass values. Furthermore, for low-mass stars, i.e. $m_i < 0.7 \, M_\odot$, the classical apsidal motion constant is taken to be $k_{\mathrm{am},i} = 0.14$ corresponding to $n=3/2$ polytropes as an estimate for these (nearly) fully convective stars. For helium (He) MS stars and WDs $k_{\mathrm{am},i}$ is calculated as function of mass by means of interpolations of tabulated data from \citet{vila77}. Lastly, for neutron stars $k_{\mathrm{am},i}$ is calculated as function of mass and radius from the expression given in \citet{hind08} and for black holes $k_{\mathrm{am},i} = 0$ as appropriate for non-spinning black holes.

\section{Triple evolution examples}
\label{sect:extrev}
To demonstrate the triple algorithm presented in Sect. \ref{sect:triplealg} we describe in detail the evolution of two triple systems in which an inner binary CO WD merger occurs in either a tight circular orbit or a wide and highly eccentric orbit. The initial conditions for these two systems are selected from our first triple population synthesis sample (TSM1, cf. Sect. \ref{sect:popsyn:samplmeth}). 

\subsection{Merger in a tight circular orbit}
\label{sect:extrev:TICM}
We consider a triple system with initial parameters $m_1 = 3.95 \, M_\odot$, $m_2 = 3.03 \, M_\odot$, $m_3 = 2.73 \, M_\odot$, $a_1 = 19.7 \, \mathrm{AU}$, $a_2 = 636.1 \, \mathrm{AU}$, $e_1 = 0.23$, $e_2 = 0.82$, $i_\mathrm{tot} = 116.0^\circ$, $g_1 = 28.5^\circ$ and $g_2 = 249.6^\circ$. Fig. \ref{fig:ex10131} shows $a_1$, $e_1$, $\beta \equiv a_2/a_1$, the tidal strength quantity $k_{\mathrm{am},i}/T_i$ \citep{hut81,hur02} and $i_\mathrm{tot}$ as a function of time. The evolution is shown with both the triple algorithm enabled (triple case; solid line) and disabled (binary case; dashed line), leaving all inner binary evolution to {\tt Binary\_c}. The latter binary case is equivalent to the situation in which no tertiary is present. 

In the binary case the semi-major axis increases at the two moments when the binary stars evolve from AGB stars to CO WDs (at $t \approx 0.2 \, \mathrm{Gyr}$ and $t \approx 0.5 \, \mathrm{Gyr}$ respectively) as a consequence of the associated wind mass loss. Here (i.e. in {\tt Binary\_c}) it is assumed that the wind mass loss is fast and isotropic such that $a_1(m_1+m_2)$ remains approximately constant. When either star in the inner binary system enters the RGB or AGB phase it develops a convective envelope and its radius increases significantly. The change of the envelope structure is reflected by a substantial increase of the tidal strength quantities $k_{\mathrm{am},i}/T_i$ as is shown in Fig. \ref{fig:ex10131} (primary AGB phase).  However, despite the large $k_{\mathrm{am},1}/T_1$ and primary radius there is no significant tidal friction during the primary AGB phase because of the large $a_1$ and small (and constant) $e_1$. Consequently, in the binary case the CO WDs end their evolution in a wide and non-interacting binary with $a_1 \approx 84 \, \mathrm{AU}$. 

\begin{figure*}
\center
\includegraphics[scale = 1.4, trim = 0mm 1mm 0mm 0mm]{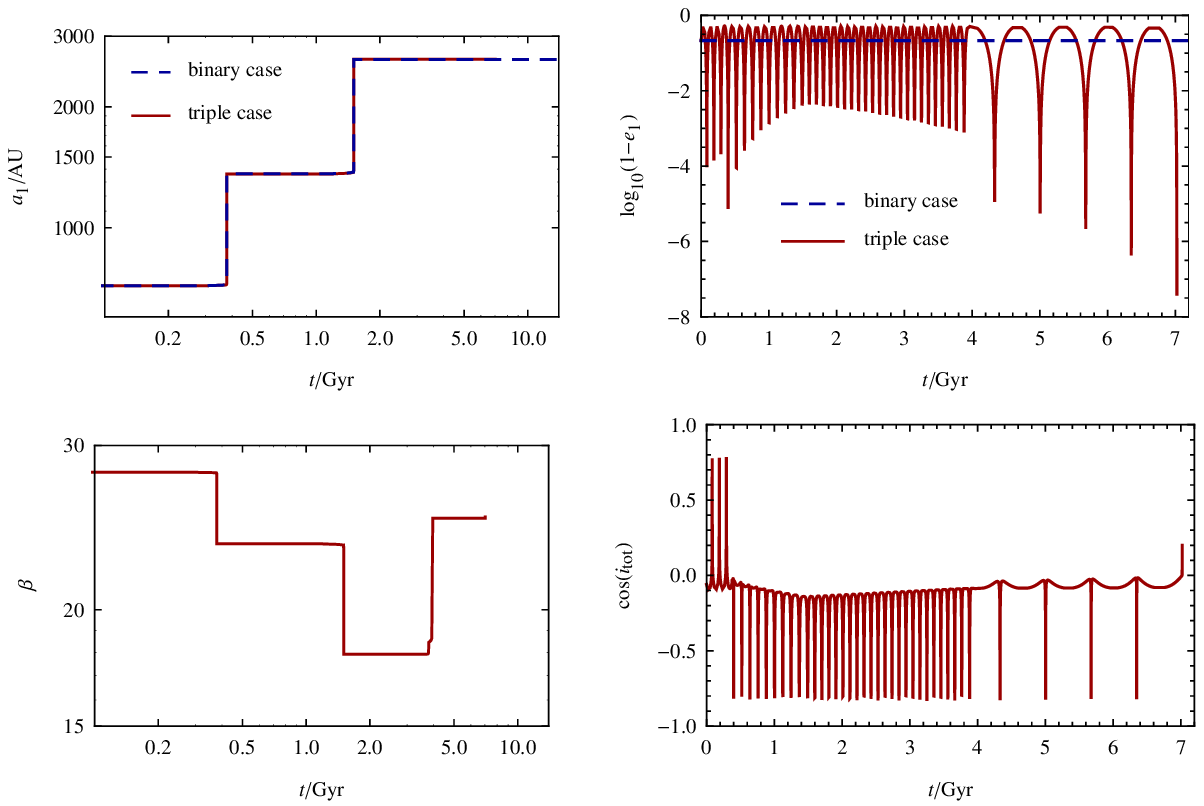}
\caption{\small Quantities of interest for the second example system in which an inner binary CO WD merger occurs in a wide and highly eccentric orbit. }
\label{fig:ex42406}
\end{figure*}

In the triple case high-amplitude Kozai cycles are induced during the MS in the inner orbit by the tertiary with maxima of $e_{1,\mathrm{max}} \approx 0.9$. Octupole order terms are important because $|\epsilon_\mathrm{oct}| \approx 10^{-2}$. Nevertheless,  during the MS the eccentricity is not high enough to drive significant tidal friction because the tidal strength quantity is very small ($k_{\mathrm{am},i}/T_i \sim 10^{-18} \, \mathrm{s^{-1}}$) as a consequence of the radiative envelopes in the MS stars. During the primary RGB phase the primary possesses a  convective envelope ($k_{\mathrm{am},1}/T_1 \sim 10^{-8} \, \mathrm{s^{-1}}$) but $e_1$ is not high enough to trigger significant tidal friction. However, tidal friction does become effective during the primary AGB phase starting at $t \approx 220.6 \, \mathrm{Myr}$ and circularizes the inner orbit during the time span of five Kozai cycles, where significant orbital shrinkage occurs at eccentricity maxima. Note that during this phase the increase in the effectiveness of orbital shrinkage is due to the expansion of the primary from a radius of $R_1 \approx 49 \, R_\odot$ to $R_1 \approx 497 \, R_\odot$ between $t = 220 \, \mathrm{Myr}$ and $t = 222.5 \, \mathrm{Myr}$. Consequently $a_1$ is reduced to $a_1 \approx 7 \, \mathrm{AU}$ and the orbit is completely circularized. Note that for complete circularization to occur the duration of the phase in which $k_{\mathrm{am},1}/T_1$ is substantial must be sufficiently long compared to the Kozai period $P_K$ (Eq. \ref{eq:PK}), which is the case for this example system. In other cases of triple systems, however, $P_K$ can be much longer than the duration of the RGB/AGB phases, thus avoiding strong tidal friction even if the eccentricity maxima are high.

Shortly after KCTF during the primary AGB phase the primary has swelled up to $R_1 \approx 611 \, R_\odot$, fills its Roche-lobe and invokes CE evolution, shrinking the orbit further to $a_1 \approx 1 \, \mathrm{AU}$ ($t \approx 222.6 \, \mathrm{Myr}$). During this process the ratio $\beta$ increases substantially to $\beta \approx 1321$, which is large enough for general relativistic precession to fully quench any subsequent Kozai cycles. Consequently, the inclination angle $i_\mathrm{tot}$ is frozen to $i_\mathrm{tot} \approx 128^\circ$. The primary emerges from the CE as a CO WD and the secondary  as a MS star. At $t \approx 0.5 \, \mathrm{Gyr}$ the secondary evolves to an AGB star itself, invoking a second CE phase. A close CO WD binary with $a_1 \approx 0.01 \, \mathrm{AU}$ emerges from the CE. Due to GW emission the system subsequently merges at $t \approx 13.2 \, \mathrm{Gyr}$ in a circular orbit. The combined CO WD mass, $1.45 \, M_\odot$, exceeds the Chandrasekhar mass $M_\mathrm{Ch} \approx 1.4 \, M_\odot$, therefore the merger could potentially lead to a SN Ia explosion (cf. Sect. \ref{sect:implications:likelihood}). 

\subsection{Merger in a wide eccentric orbit}
\label{sect:extrev:TIEM}
The second example system has initial parameters $m_1 = 3.26 \, M_\odot$, $m_2 = 2.00 \, M_\odot$, $m_3 = 1.39 \, M_\odot$, $a_1 = 716.6 \, \mathrm{AU}$, $a_2 = 2.012 \cdot 10^{4} \, \mathrm{AU}$, $e_1 = 0.78$, $e_2 = 0.64$, $i_\mathrm{tot} = 93.4^\circ$, $g_1 = 150.0^\circ$ and $g_2 = 151.2^\circ$. The evolution of $a_1$, $e_1$, $\beta \equiv a_2/a_1$ and $i_\mathrm{tot}$ for this system as function of time is shown in Fig. \ref{fig:ex42406}. The main difference in initial parameters in this example system compared to the previous example is that $a_1$ and $a_2$ are larger. Their ratio $\beta$ is comparable, however. Because of the larger $a_2$ the Kozai period $P_K \approx  100 \, \mathrm{Myr}$ is much longer than in the first example ($P_K \approx 0.3 \, \mathrm{Myr}$). Consequently, it is much more unlikely that the primary RGB and AGB phases (which occur at $t \approx 0.3 \, \mathrm{Gyr}$ and $t \approx 0.4 \, \mathrm{Gyr}$ respectively) coincide with eccentricity maxima. Such a coincidence is indeed not the case in the second example system.

As the primary and secondary evolve to CO WDs without interacting the associated wind mass loss widens the orbit to $a_1 \approx 2600 \, \mathrm{AU}$. Although this inner binary wind mass loss increases $a_2$ as well, the ratio $\beta = a_2/a_1$ decreases with the assumption of fast isotropic winds such that $a_1(m_1+m_2)$ and $a_2(m_1+m_2+m_3)$ are constant (in this example $m_3$ also remains constant). Consequently, the ratio $\beta$ decreases to $\beta \approx 18$ after the secondary has evolved to a CO WD. As the tertiary evolves to a CO WD and subsequently loses mass at $t \approx 1.5 \, \mathrm{Gyr}$ $a_2$ increases while $a_1$ stays constant, thus increasing $\beta$ again. The latter process also increases the Kozai period. After $t \approx 1.5 \, \mathrm{Gyr}$ the maximum values of $\cos(i_\mathrm{tot})$, which are still negative (retrograde orbit), slowly increase and the corresponding eccentricity maxima increase as well. As $\cos(i_\mathrm{tot})$ approaches 0 the eccentricity maxima reach very high values until at $t \approx 7.0 \, \mathrm{Gyr}$ the orbits become prograde, which is associated with $e_1$ reaching a value of $1-e_1 \approx 10^{-7.4}$. This is high enough to trigger an orbital collision between the two white dwarfs, even though $a_1 \approx 2600 \, \mathrm{AU}$ is very large. Such a collision could potentially result in a SN Ia explosion as in the previous example (cf. Sect. \ref{sect:implications:likelihood}), but note that the nature of the merger prior to a possible SNe Ia event is very different. 

One might wonder whether in this system tidal friction is significant during close periastron passages before the collision occurs. For this reason we have recalculated this system with the tidal strength quantities $k_{\mathrm{am},i}/T_i$ for radiative and degenerate damping multiplied by ad hoc factors of $10^3$ and $10^6$ to artificially increase the effectiveness of tidal friction. Even with these multiplications we find that the outcome is identical, showing that tidal friction is not important in this system. We have performed similar tests for all systems in the population synthesis study and have found no effect (see Sect. \ref{sect:discussion:uncertainties}).

\section{Population synthesis: methods}
\label{sect:methods}

\begin{figure}
\center
\includegraphics[scale = 0.8, trim = 0mm 0mm 0mm 0mm]{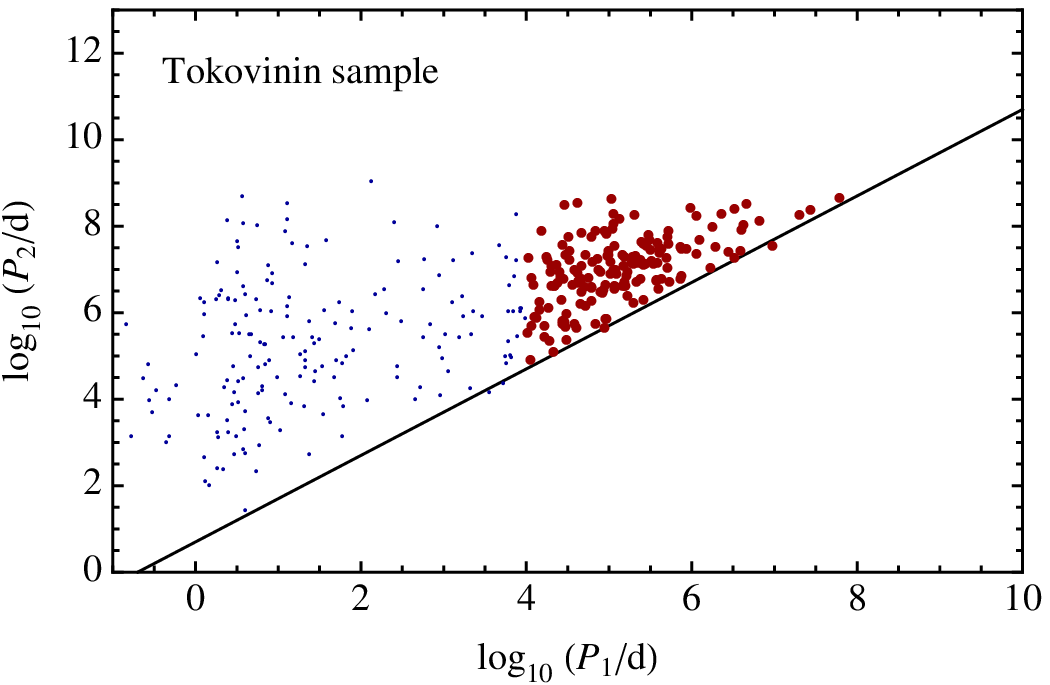}
\includegraphics[scale = 0.8, trim = 0mm 0mm 0mm 0mm]{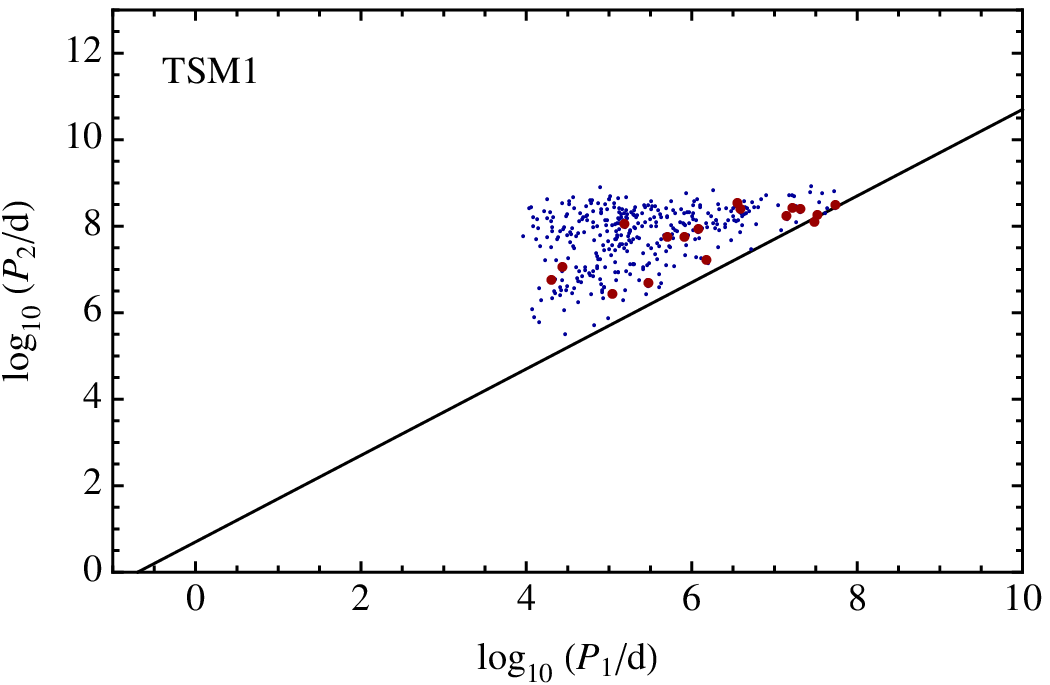}
\includegraphics[scale = 0.8, trim = 0mm 0mm 0mm 0mm]{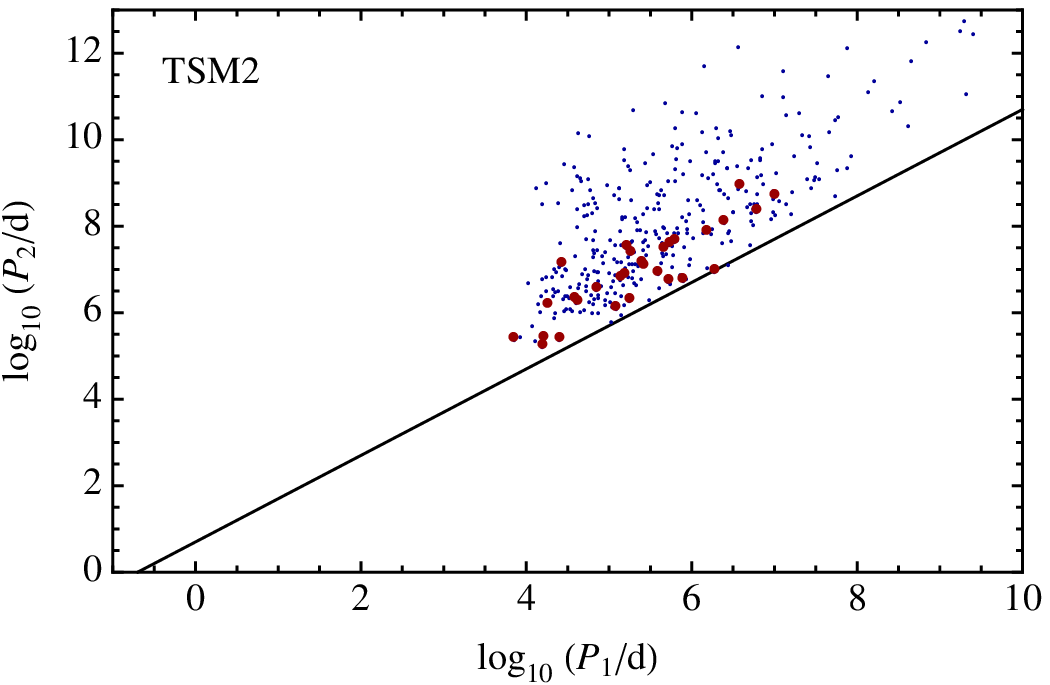}
\caption{\small Inner ($P_1$) and outer ($P_2$) orbital periods of triple systems of the observed triple sample of \citet{tok08} (top) and the two synthetic triple samples described in Sect. \ref{sect:popsyn:samplmeth} (middle and bottom). In the observational sample systems satisfying our selection criteria are denoted with large (red) dots ({\color{red} \textbullet}). In the synthetic samples the large dots ({\color{red} \textbullet}) denote the systems that undergo an early MS inner binary merger or destabilization and are therefore not likely to be observed. The line corresponds to $P_2/P_1 = 5$ and represents the typical critical value for dynamical stability. }
\label{fig:p1p2diag}
\end{figure}

Before presenting the results of our population synthesis study (Sect. \ref{sect:results}) we discuss here our selection criteria and sampling methods which are essential ingredients for such a study. 

\subsection{Selection criteria}
\label{sect:popsyn:selcrit}
Because we are mainly interested in inner binary mergers involving a CO WD we take an upper limit of the inner binary primary mass $m_1$ of $m_{1,u} = 6.5 \, M_\odot$, which approximately captures the boundary between forming a CO WD and an ONe WD in the case of single star evolution. The lower limit of $m_1$ is set to $m_{1,l} = 1.0 \, M_\odot$ for any lower mass will not produce a significant number of CO WDs within a Hubble time for a metallicity of $Z = 0.02$. In addition we focus on triple systems with wide inner binary systems such that the inner binary stars do not interact during their evolution in the absence of a tertiary. Taking into account possible tidal friction this can be achieved by requiring that the initial inner orbit semi-latus rectum $l_1 \equiv a_1 \left(1-e_1^2\right)$ satisfies $l_1 > l_{1,l} \equiv 12 \, \mathrm{AU}$.

One possible method to obtain the initial triple parameters is to directly sample data from observed triple systems. Fig. \ref{fig:p1p2diag} (top) shows the observed sample of 725 triple systems of \citet{tok08} in the $(P_1,P_2)$-diagram. A clear paucity of systems with inner orbital periods between $\log_{10}(P_1/\mathrm{d}) = 2$ and $\log_{10}(P_1/\mathrm{d}) = 4$ is present, which is likely because of the difficulty of detecting such systems in the regime between spectroscopic and visual binaries. Our selection criteria of inner binary systems with $1.0 < m_1/M_\odot < 6.5$ and $l_1 > 12 \, \mathrm{AU}$ decrease the number of systems from 725 to 165, where a value of $\log_{10}(P_1/\mathrm{d}) = 4$ is taken to represent the dividing value $l_{1,l} = 12 \, \mathrm{AU}$. The number of remaining systems (indicated in the top figure of Fig. \ref{fig:p1p2diag} by dots) is not large enough for direct sampling, therefore we choose to utilize the method of Monte Carlo sampling from observation-based distributions. In order to gain information on the uncertainties of our results we use two distinct sampling methods that are described below.

\subsection{Sampling methods}
\label{sect:popsyn:samplmeth}
\subsubsection{TSM1}
The first triple sampling method (TSM1) is based on the supposition that a hierarchical triple system is composed of two uncorrelated binary systems. The main advantage of this approach is that the statistics of binary parameters are known with greater certainty than those of triple systems. A primary mass $1.0<m_1/M_\odot<6.5$ is sampled from a \citet{kroupa93} initial mass function (IMF), i.e. $\mathrm{d} N/\mathrm{d} m_1 \propto m_1^{-2.70}$. Subsequently, two mass ratios, $q_1 \equiv m_2 / m_1$ and $q_2 \equiv m_3/(m_1+m_2)$, are sampled independently from a uniform distribution such that $0<q_j\leq1$, $m_2 > 0.01 \, M_\odot$ and $m_3 > 0.01 \, M_\odot$, consistent with Claeys et al. (in prep.). The lower limit on the stellar masses is in accordance with the limit found by \citet{kouw07}. The secondary and tertiary masses are subsequently computed from these sampled mass ratios. We sample $e_1$ and $e_2$ from a thermal distribution $\mathrm{d}N/\mathrm{d} e_j = 2 e_j$ as is appropriate for $P_j > 10^3 \, \mathrm{d}$ binaries \citep{kroupa01}. Two semi-major axes are then sampled from a distribution that is flat in $\log_{10} (a_j)$, the smaller of which is designated $a_1$ and the larger of which is designated $a_2$. The lower and upper limits $a_l$ and $a_u$ in the distribution of the semi-major axes are $a_l = 5 \, R_\odot$ and $a_u = 5 \cdot 10^6 \, R_\odot$ for both inner and outer orbits \citep{kouw07}. From these systems we reject those that do not satisfy $l_1 = a_1 \left( 1 - e_1^2 \right ) > l_{1,l} = 12 \, \mathrm{AU}$. Lastly, we sample the initial orbital inclination angle $i_\mathrm{tot}$ from a distribution that is uniform in $\cos(i_\mathrm{tot})$ with $ -1<\cos(i_\mathrm{tot}) < 1$ and the initial arguments of periastron of both orbits, $g_1$ and $g_2$, from a uniform distribution with $0<g_j<2 \pi$. From the triple systems obtained in this manner we subsequently reject the systems that are not dynamically stable based on the stability criterion of \citet{mard01}, consistent with our triple evolution algorithm (cf. Sect. \ref{sect:triplealg:binalg}). 

\subsubsection{TSM2}
In the second triple sampling method (TSM2) we use the multiple system recipe developed by \citet{egg09}. This recipe is designed to reproduce the properties of a set of 4558 stellar systems with Hipparcos magnitude brighter than 6 collected by \citet{eggtok08}. It is followed until the second bifurcation, effectively restricting to multiple systems that are triple. The parameters given by the recipe are $m_1$, $m_2$, $m_3$, $P_1$ and $P_2$. The primary mass distribution in this recipe is designed to resemble the Salpeter IMF ($\mathrm{d} N/\mathrm{d} m_1 \propto m_1^{-2.35}$) at large masses and turns over for masses below $0.3 \, M_\odot$. The mass ratio distribution is approximately flat and the $P_2$ distribution has a broad peak around $10^5 \, \mathrm{d}$. We refer to \citet{egg09} for further details. The corresponding semi-major axes are computed from Kepler's third law. The parameters not prescribed by the recipe are $i_\mathrm{tot}$, $e_j$ and $g_j$, which are sampled from the same distributions as in TSM1 above. Similarly to the method in TSM1 systems are rejected if $l_1 \leq l_{1,l} = 12 \, \mathrm{AU}$ and if they do not satisfy the stability criterion of \citet{mard01}.

\newcommand{\sampleTok}[2]{\pgfplotstablegetelem{#1}{#2}\of{\filesampleTok}  \pgfplotsretval}
\pgfplotstableread{sampleTok.txt}{\filesampleTok}

\newcommand{\sampleTSMone}[2]{\pgfplotstablegetelem{#1}{#2}\of{\filesampleTSMone}  \pgfplotsretval}
\pgfplotstableread{sampleTSM1.txt}{\filesampleTSMone}
\newcommand{\sampleTSMtwo}[2]{\pgfplotstablegetelem{#1}{#2}\of{\filesampleTSMtwo}  \pgfplotsretval}
\pgfplotstableread{sampleTSM2.txt}{\filesampleTSMtwo}

\begin{table}
\begin{center}
\begin{tabular}{lcccccc}
\toprule 
& \multicolumn{2}{c}{Tokovinin} & \multicolumn{2}{c}{TSM1} & \multicolumn{2}{c}{TSM2} \\
\cmidrule(r){2-3} \cmidrule(r){4-5} \cmidrule(r) {6-7}
& \multicolumn{1}{c}{{Mean}} & \multicolumn{1}{c}{SD} & \multicolumn{1}{c}{{Mean}} & \multicolumn{1}{c}{SD} & \multicolumn{1}{c}{{Mean}} & \multicolumn{1}{c}{SD} \\ 
\midrule
$m_1/M_\odot$ & \sampleTok{0}{0} & \sampleTok{0}{1} & \sampleTSMone{0}{0} & \sampleTSMone{0}{1} & \sampleTSMtwo{0}{0} & \sampleTSMtwo{0}{1} \\
$m_2/M_\odot$ & \sampleTok{1}{0} & \sampleTok{1}{1} & \sampleTSMone{1}{0} & \sampleTSMone{1}{1} & \sampleTSMtwo{1}{0} & \sampleTSMtwo{1}{1} \\
$m_3/M_\odot$ & \sampleTok{2}{0} & \sampleTok{2}{1} & \sampleTSMone{2}{0} & \sampleTSMone{2}{1} & \sampleTSMtwo{2}{0} & \sampleTSMtwo{2}{1} \\
$\log_{10} (P_1/\mathrm{d})$ & \sampleTok{3}{0} & \sampleTok{3}{1} & \sampleTSMone{3}{0} & \sampleTSMone{3}{1} & \sampleTSMtwo{3}{0} & \sampleTSMtwo{3}{1} \\
$\log_{10} (P_2/\mathrm{d})$ & \sampleTok{4}{0} & \sampleTok{4}{1} & \sampleTSMone{4}{0} & \sampleTSMone{4}{1} & \sampleTSMtwo{4}{0} & \sampleTSMtwo{4}{1} \\
\bottomrule
\end{tabular}
\end{center}
\caption{\small Mean and standard deviation (SD) of masses and orbital periods of the \citet{tok08} sample satisfying the selection criteria (Sect. \ref{sect:popsyn:selcrit}) and of both sampled triple populations TSM1 and TSM2. }
\label{table:TSMstat}
\end{table}
 
\subsubsection{Properties and comparisons}
Here we elaborate on some of the properties of the sampled and observed populations. Fig. \ref{fig:p1p2diag} (middle and bottom) shows the $(P_1,P_2)$-diagram for a small sample (360 systems) of the TSM1 and TSM2 populations. The upper boundaries of $P_1$ and $P_2$ in both populations are quite different. In TSM1 $P_1$ and $P_2$ are limited by $a_l$ and $a_u$, whereas in TSM2 such sharp boundaries do not exist. In TSM2 the upper boundary of $P_2$ is dependent on $P_1$ because in this sampling method the largest ratio $P_2/P_1$ is given by $P_2/P_1 = 2 \cdot 10^6$. These wide outer binary systems are very rare, however. Fig. \ref{fig:p1p2diag} (middle and bottom) also shows the systems that according to our triple algorithm experience a merger in the inner binary system or a destabilization early on the MS (cf. Sects. \ref{sect:results:mergers:MS} and \ref{sect:results:destab}). Such systems are not likely part of an observed sample of triples because of their short lifetimes. Because of their small number, removing these systems does not significantly affect the distributions of masses and orbital periods (the mean values are affected by typically 1\%).

In Table \ref{table:TSMstat} the distributions of the masses and orbital periods of the Tokovinin sample with our selection criteria applied are compared to the two sampled populations. Because of the steep slope of $-2.70$ in the TSM1 IMF the mean masses in TSM1 are small compared to those in TSM2, for which a Salpeter-like slope (-2.35) applies for $m_1 > 1.0 \, M_\odot$. Compared to the Tokovinin sample the masses of TSM1 and TSM2 appear to be on the low side, but it is important to note that the Tokovinin sample is magnitude-limited rather than volume-limited, therefore it is strongly biased towards more massive systems. The distributions of the orbital periods of TSM1 and TSM2 are similar, the orbital periods being on average somewhat higher in TSM2 compared to those in TSM1. The orbital periods of the Tokovinin sample, notably the outer orbital periods, are smaller than those in the sampled populations. This is also likely affected by observational bias, because very long orbital periods on the order of $10^7$ days or longer are generally very difficult to measure. 

\begin{figure}
\center
\includegraphics[scale = 0.78, trim = 0mm 0mm 0mm 0mm]{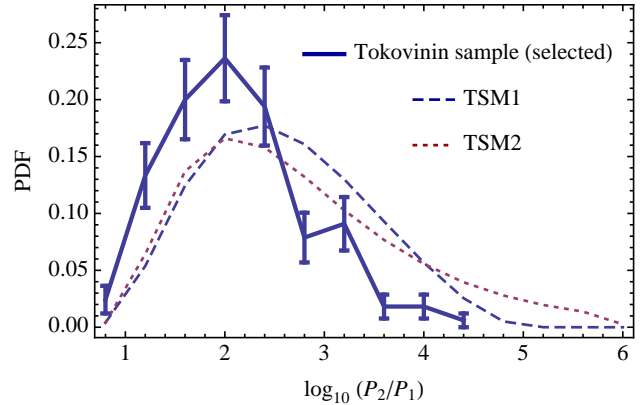}
\caption{\small Probability density functions (PDFs) of $P_2/P_1$ for the Tokovinin sample with our selection criteria applied (solid line), the TSM1 sample (dashed line) and the TSM2 sample (dotted line). For the Tokovinin sample, errors based on Poisson statistics are indicated.}
\label{fig:p1p2comp}
\end{figure}

Lastly, Fig. \ref{fig:p1p2comp} compares the distributions of the ratio $P_2/P_1$ for the three populations. For TSM1 and TSM2 these distributions are similar. The distribution of $P_2/P_1$ of the Tokovinin sample is peaked towards smaller ratios, although here observational bias for large period ratios may be important. Note also the large error bars in the Tokovinin sample as a consequence of the rather small number of systems in the Tokovinin sample that satisfy our selection criteria (165).

\subsubsection{Sampling procedure}
We sample $N_\mathrm{calc} = 2 \cdot 10^6$ triple systems for both populations TSM1 and TSM2. Using the triple algorithm described in Sect. \ref{sect:triplealg} these systems are evolved from $t = 0$, with all three components starting as ZAMS stars, to a Hubble time $t_H = 13.7 \, \mathrm{Gyr}$. In order to retain sufficient resolution in higher-mass systems ($2.0 < m_1/M_\odot < 6.5$) we split both TSM1 and TSM2 into two parts of $1 \cdot 10^6$ systems, one with $1.0 < m_1/M_\odot < 2.0$ and one with $2.0 < m_1/M_\odot < 6.5$. In the results presented in Sect \ref{sect:results} systems in each part are given appropriate weights determined by the mass function of $m_1$ to account for the fact that the actual number of systems occurring in the $1.0 < m_1/M_\odot < 2.0$ range is larger than that in the $2.0 < m_1/M_\odot < 6.5$ range (see Appendix \ref{app:DTDcalc} for details). Several key events in the evolution are kept track of, most notably the onset of significant KCTF, CE evolution, a merger in the inner binary system and a destabilization of the triple system. We thus obtain a catalogue of the evolutionary outcomes where we distinguish between three main channels: inner binary mergers, no inner binary mergers and triple destabilizations. In a very small number of systems (55 and 44 for TSM1 and TSM2 respectively) the calculation of the evolution cannot be completed due to convergence errors in the ODE solver routine (cf. Sect. \ref{sect:triplealg}); these systems are excluded from the results below. In these cases the errors occur just prior to a MS merger or RGB + MS merger. Because the affected systems constitute a tiny fraction of all systems for which a MS merger or RGB + MS merger applies this does not affect our conclusions.

\begin{table}
\begin{center}
\begin{tabular}{cl}
\toprule 
$k$ & Description \\
\midrule
0 & MS ($m\lesssim0.7 \, M_\odot$) \\
1 & MS ($m\gtrsim0.7 \, M_\odot$) \\
2 & Hertzsprung gap (HG) \\
3 & red giant branch (RGB) \\
4 & core helium burning (CHeB) \\
5 & early asymptotic giant branch (EAGB) \\
6 & thermally pulsing AGB (TPAGB) \\
7 & naked helium star MS (He MS) \\
8 & naked helium star Hertzsprung gap (He HG) \\
9 & naked helium star giant branch (He GB) \\
10 & helium white dwarf (He WD) \\
11 & carbon-oxygen white dwarf (CO WD) \\
12 & oxygen-neon white dwarf (ONe WD) \\
13 & neutron star (NS) \\
14 & black hole (BH) \\
15 & massless remnant \\
\bottomrule
\end{tabular}
\end{center}
\caption{\small Description of the different stellar types used in Fig. \ref{figfrac}. Identical to the types used in \citet{hur02}. }
\label{table:stdes}
\end{table}

\begin{figure*}
\center
\includegraphics[scale = 0.266, trim = 0mm 15mm 0mm 0mm]{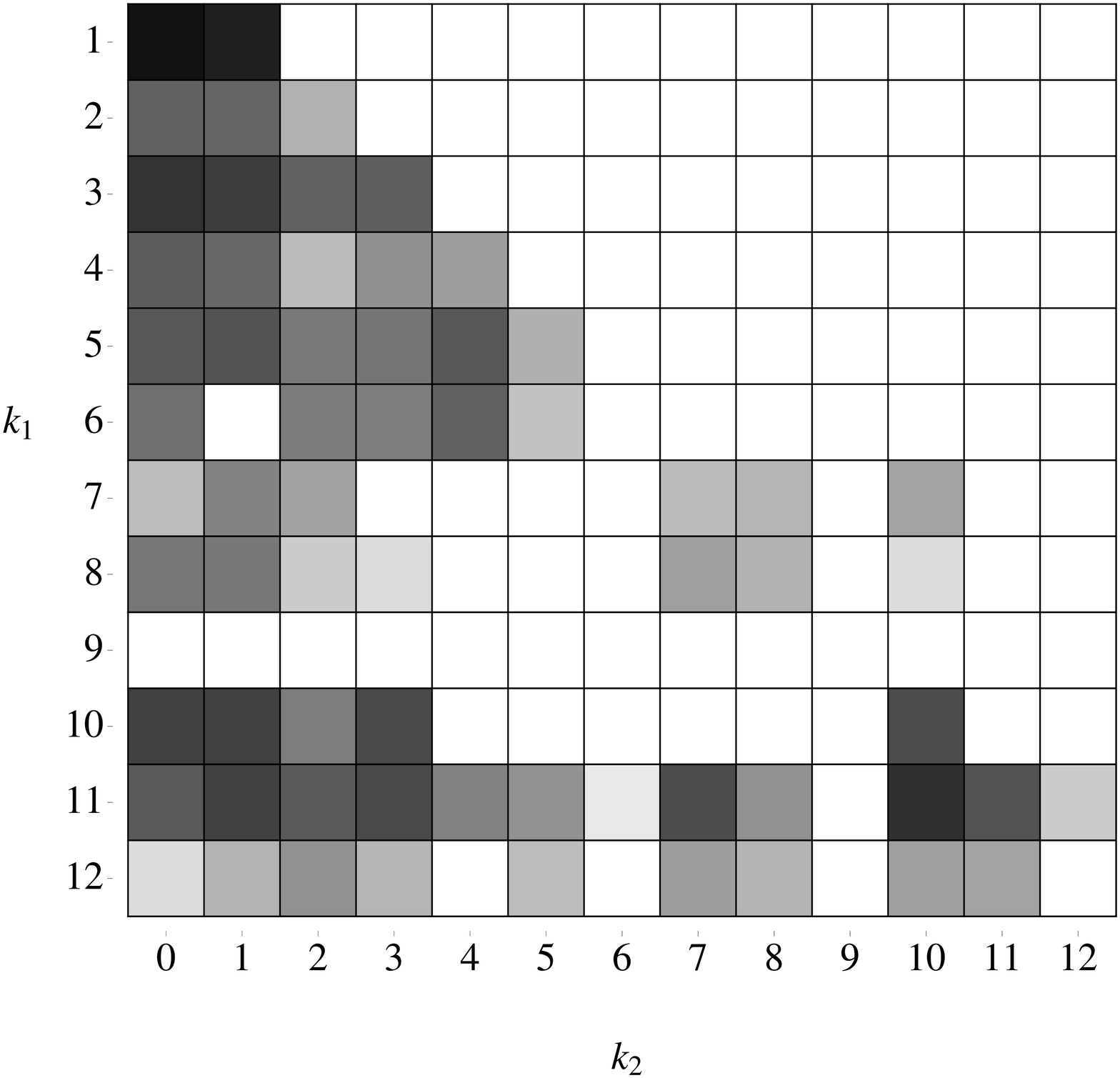}
\includegraphics[scale = 0.278, trim = 0mm 15mm 0mm 0mm]{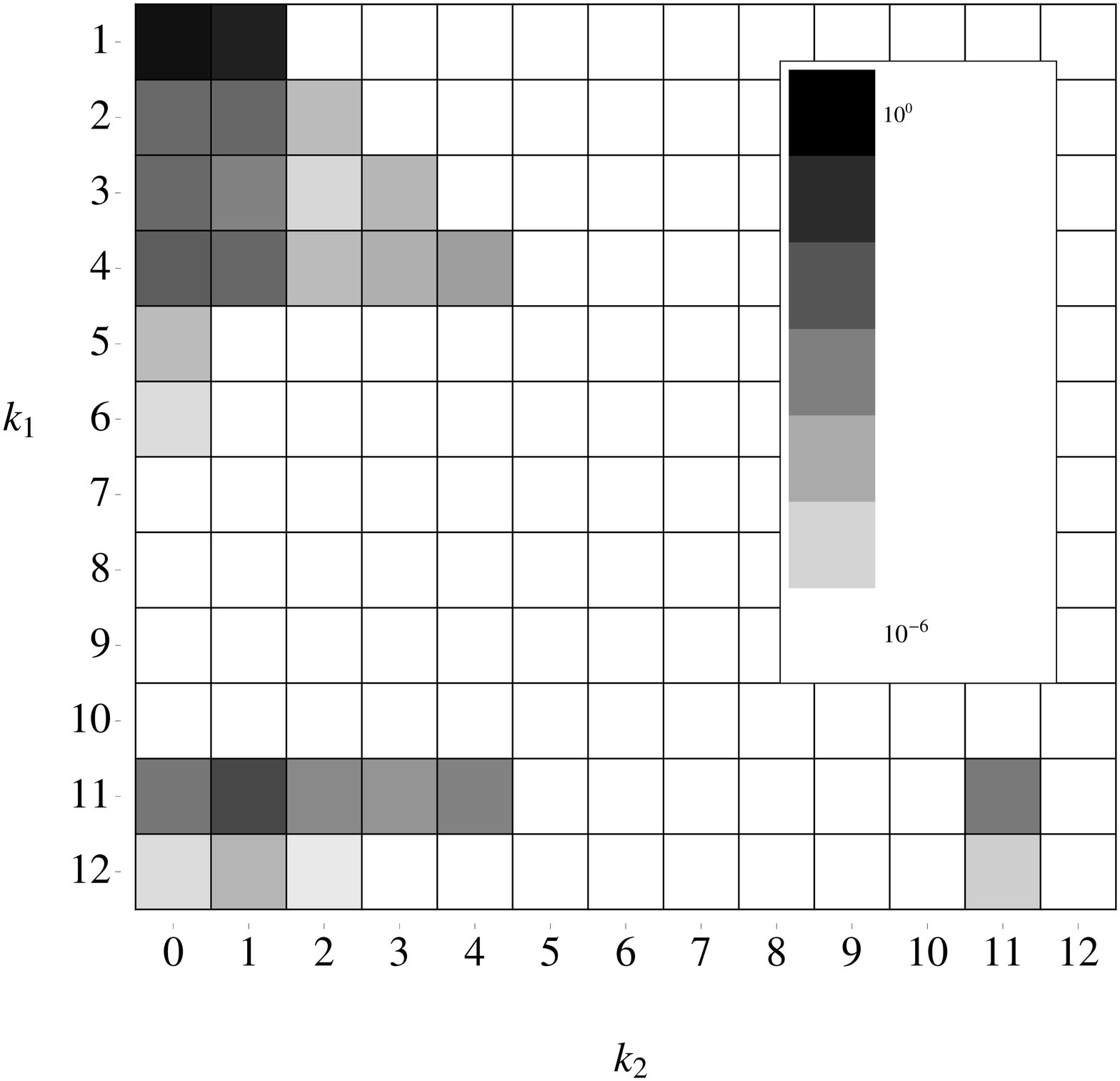}
\caption{\small Left: the number of systems in TSM1 that experience an inner binary merger for each occurring combination of stellar types just prior to merging, normalized to the total number of merger systems (that constitute $\approx 8 \%$ of all sampled triple systems). Right: the number of systems in TSM1 in which the merger occurs due to an eccentric collision, again normalized to the total number of merger systems. The stellar types are identical to the ones used in \citet{hur02} and are repeated for convenience in Table \ref{table:stdes}.}
\label{figfrac}
\end{figure*}

\section{Population synthesis: results}
\label{sect:results}

\newcommand{\maintable}[2]{\pgfplotstablegetelem{#1}{#2}\of{\filemaintable}  \pgfplotsretval}
\pgfplotstableread{maintable_zeros.txt}{\filemaintable}

\renewcommand{\tabcolsep}{0.11cm}

\begin{table*}
\begin{tabular}{lrrcccccccccccccc}
\toprule 
& & & & \multicolumn{10}{c}{$f_\mathrm{KCTF}$} \\
\cmidrule(r){4-13}
& \multicolumn{2}{c}{$P$} & \multicolumn{2}{c}{MS} & \multicolumn{2}{c}{PHG/PRGB} & \multicolumn{2}{c}{PAGB} & \multicolumn{2}{c}{SHG/SRGB} & \multicolumn{2}{c}{SAGB} & \multicolumn{2}{c}{$f_\mathrm{CE}$} & \multicolumn{2}{c}{$f_\mathrm{EC}$} \\ 
\cmidrule(r){2-3} \cmidrule(r){4-5} \cmidrule(r){6-7} \cmidrule(r){8-9} \cmidrule(r){10-11} \cmidrule(r){12-13} \cmidrule(r){14-15} \cmidrule(r){16-17}
& TSM1 & TSM2 & TSM1 & TSM2 & TSM1 & TSM2 & TSM1 & TSM2 & TSM1 & TSM2 & TSM1 & TSM2 & TSM1 & TSM2 & TSM1 & TSM2 \\
\midrule
Merger & {\bf \maintable{0}{0}} & {\bf \maintable{0}{1}} & \maintable{0}{2} & \maintable{0}{3} & \maintable{0}{4} & \maintable{0}{5} & \maintable{0}{6} & \maintable{0}{7} & \maintable{0}{8} & \maintable{0}{9} & \maintable{0}{10} & \maintable{0}{11} & \maintable{0}{12} & \maintable{0}{13} & \maintable{0}{14} & \maintable{0}{15} \\
\midrule[0.08pt]
$\bullet$ On MS $\rightarrow$ single MS & {\bf \maintable{1}{0}} & {\bf \maintable{1}{1}} & \maintable{1}{2} & \maintable{1}{3} & --- & --- & --- & --- & --- & --- & --- & --- & \maintable{1}{12} & \maintable{1}{13} & \maintable{1}{14} & \maintable{1}{15} \\
\midrule[0.08pt]
$\quad$ $\rightarrow$ CO WD & \maintable{2}{0} & \maintable{2}{1} & \maintable{2}{2} & \maintable{2}{3} & --- & --- & --- & --- & --- & --- & --- & --- & \maintable{2}{12} & \maintable{2}{13} & \maintable{2}{14} & \maintable{2}{15} \\
$\quad$ $\rightarrow$ ONe WD & \maintable{3}{0} & \maintable{3}{1} & \maintable{3}{2} & \maintable{3}{3} & --- & --- & --- & --- & --- & --- & --- & --- & \maintable{3}{12} & \maintable{3}{13} & \maintable{3}{14} & \maintable{3}{15} \\
$\quad$ $\rightarrow$ CC SN $\rightarrow$ NS & \maintable{4}{0} & \maintable{4}{1} & \maintable{4}{2} & \maintable{4}{3} & --- & --- & --- & --- & --- & --- & --- & --- & \maintable{4}{12} & \maintable{4}{13} & \maintable{4}{14} & \maintable{4}{15} \\
\midrule[0.08pt]
$\bullet$ Post-MS & {\bf \maintable{5}{0}} & {\bf \maintable{5}{1}} & \maintable{5}{2} & \maintable{5}{3} & \maintable{5}{4} & \maintable{5}{5} & \maintable{5}{6} & \maintable{5}{7} & \maintable{5}{8} & \maintable{5}{9} & \maintable{5}{10} & \maintable{5}{11} & \maintable{5}{12} & \maintable{5}{13} & \maintable{5}{14} & \maintable{5}{15} \\
\midrule[0.08pt]
$\quad$ $\bullet$ HG + MS (2) & \maintable{6}{0} & \maintable{6}{1} & \maintable{6}{2} & \maintable{6}{3} & \maintable{6}{4} & \maintable{6}{5} & --- & --- & --- & --- & --- & --- & \maintable{6}{12} & \maintable{6}{13} & \maintable{6}{14} & \maintable{6}{15} \\
$\quad$ $\bullet$ RGB + MS (3) & \maintable{7}{0} & \maintable{7}{1} & \maintable{7}{2} & \maintable{7}{3} & \maintable{7}{4} & \maintable{7}{5} & --- & --- & --- & --- & --- & --- & \maintable{7}{12} & \maintable{7}{13} & \maintable{7}{14} & \maintable{7}{15} \\
$\quad$ $\bullet$ CHeB + MS (4) & \maintable{8}{0} & \maintable{8}{1} & \maintable{8}{2} & \maintable{8}{3} & \maintable{8}{4} & \maintable{8}{5} & --- & --- & --- & --- & --- & --- & \maintable{8}{12} & \maintable{8}{13} & \maintable{8}{14} & \maintable{8}{15} \\
$\quad$ $\bullet$ AGB + MS (5) & \maintable{9}{0} & \maintable{9}{1} & \maintable{9}{2} & \maintable{9}{3} & \maintable{9}{4} & \maintable{9}{5} & \maintable{9}{6} & \maintable{9}{7} & --- & --- & --- & --- & \maintable{9}{12} & \maintable{9}{13} & \maintable{9}{14} & \maintable{9}{15} \\
$\quad$ $\bullet$ AGB + CHeB (4) & \maintable{10}{0} & \maintable{10}{1} & \maintable{10}{2} & \maintable{10}{3} & \maintable{10}{4} & \maintable{10}{5} & \maintable{10}{6} & \maintable{10}{7} & \maintable{10}{8} & \maintable{10}{9} & --- & --- & \maintable{10}{12} & \maintable{10}{13} & \maintable{10}{14} & \maintable{10}{15} \\
\midrule[0.08pt]
$\bullet$ Compact object & {\bf \maintable{11}{0}} & {\bf \maintable{11}{1}} & \maintable{11}{2} & \maintable{11}{3} & \maintable{11}{4} & \maintable{11}{5} & \maintable{11}{6} & \maintable{11}{7} & \maintable{11}{8} & \maintable{11}{9} & \maintable{11}{10} & \maintable{11}{11} & \maintable{11}{12} & \maintable{11}{13} & \maintable{11}{14} & \maintable{11}{15} \\
\midrule[0.08pt]
$\quad$ $\bullet$ He WD + MS (3) & \maintable{12}{0} & \maintable{12}{1} & \maintable{12}{2} & \maintable{12}{3} & \maintable{12}{4} & \maintable{12}{5} & \maintable{12}{6} & \maintable{12}{7} & --- & --- & --- & --- & \maintable{12}{12} & \maintable{12}{13} & \maintable{12}{14} & \maintable{12}{15} \\
$\quad$ $\bullet$ He WD + RGB (7) & \maintable{13}{0} & \maintable{13}{1} & \maintable{13}{2} & \maintable{13}{3} & \maintable{13}{4} & \maintable{13}{5} & \maintable{13}{6} & \maintable{13}{7} & \maintable{13}{8} & \maintable{13}{9} & --- & --- & \maintable{13}{12} & \maintable{13}{13} & \maintable{13}{14} & \maintable{13}{15} \\
$\quad$ $\bullet$ He WD + He WD (7) & \maintable{14}{0} & \maintable{14}{1} & \maintable{14}{2} & \maintable{14}{3} & \maintable{14}{4} & \maintable{14}{5} & \maintable{14}{6} & \maintable{14}{7} & \maintable{14}{8} & \maintable{14}{9} & \maintable{14}{10} & \maintable{14}{11} & \maintable{14}{12} & \maintable{14}{13} & \maintable{14}{14} & \maintable{14}{15} \\
$\quad$ $\bullet$ CO WD + MS (5) & \maintable{15}{0} & \maintable{15}{1} & \maintable{15}{2} & \maintable{15}{3} & \maintable{15}{4} & \maintable{15}{5} & \maintable{15}{6} & \maintable{15}{7} & --- & --- & --- & --- & \maintable{15}{12} & \maintable{15}{13} & \maintable{15}{14} & \maintable{15}{15} \\
$\quad$ $\bullet$ CO WD + HG (5) & \maintable{16}{0} & \maintable{16}{1} & \maintable{16}{2} & \maintable{16}{3} & \maintable{16}{4} & \maintable{16}{5} & \maintable{16}{6} & \maintable{16}{7} & \maintable{16}{8} & \maintable{16}{9} & --- & --- & \maintable{16}{12} & \maintable{16}{13} & \maintable{16}{14} & \maintable{16}{15} \\
$\quad$ $\bullet$ CO WD + RGB (5) & \maintable{17}{0} & \maintable{17}{1} & \maintable{17}{2} & \maintable{17}{3} & \maintable{17}{4} & \maintable{17}{5} & \maintable{17}{6} & \maintable{17}{7} & \maintable{17}{8} & \maintable{17}{9} & --- & --- & \maintable{17}{12} & \maintable{17}{13} & \maintable{17}{14} & \maintable{17}{15} \\
$\quad$ $\bullet$ CO WD + CHeB (4) & \maintable{18}{0} & \maintable{18}{1} & \maintable{18}{2} & \maintable{18}{3} & \maintable{18}{4} & \maintable{18}{5} & \maintable{18}{6} & \maintable{18}{7} & \maintable{18}{8} & \maintable{18}{9} & --- & --- & \maintable{18}{12} & \maintable{18}{13} & \maintable{18}{14} & \maintable{18}{15} \\
$\quad$ $\bullet$ CO WD + HeMS (8) & \maintable{19}{0} & \maintable{19}{1} & \maintable{19}{2} & \maintable{19}{3} & \maintable{19}{4} & \maintable{19}{5} & \maintable{19}{6} & \maintable{19}{7} & \maintable{19}{8} & \maintable{19}{9} & \maintable{19}{10} & \maintable{19}{11} & \maintable{19}{12} & \maintable{19}{13} & \maintable{19}{14} & \maintable{19}{15} \\
$\quad$ $\bullet$ CO WD + He WD (8) & \maintable{20}{0} & \maintable{20}{1} & \maintable{20}{2} & \maintable{20}{3} & \maintable{20}{4} & \maintable{20}{5} & \maintable{20}{6} & \maintable{20}{7} & \maintable{20}{8} & \maintable{20}{9} & \maintable{20}{10} & \maintable{20}{11} & \maintable{20}{12} & \maintable{20}{13} & \maintable{20}{14} & \maintable{20}{15} \\
 $\quad$ $\bullet$ CO WD + CO WD & \maintable{21}{0} & \maintable{21}{1} & \maintable{21}{2} & \maintable{21}{3} & \maintable{21}{4} & \maintable{21}{5} & \maintable{21}{6} & \maintable{21}{7} & \maintable{21}{8} & \maintable{21}{9} & \maintable{21}{10} & \maintable{21}{11} & \maintable{21}{12} & \maintable{21}{13} & \maintable{21}{14} & \maintable{21}{15} \\
\midrule
No merger & {\bf \maintable{22}{0}} & {\bf \maintable{22}{1}} & \maintable{22}{2} & \maintable{22}{3} & \maintable{22}{4} & \maintable{22}{5} & \maintable{22}{6} & \maintable{22}{7} & \maintable{22}{8} & \maintable{22}{9} & \maintable{22}{10} & \maintable{22}{11} & \maintable{22}{12} & \maintable{22}{13} & --- & --- \\
\midrule[0.08pt]
$\quad$ $\bullet$ $a_{1,f} / \mathrm{AU} > 12$ & \maintable{23}{0} & \maintable{23}{1} & \maintable{23}{2} & \maintable{23}{3} & \maintable{23}{4} & \maintable{23}{5} & \maintable{23}{6} & \maintable{23}{7} & \maintable{23}{8} & \maintable{23}{9} & \maintable{23}{10} & \maintable{23}{11} & \maintable{23}{12} & \maintable{23}{13} & --- & --- \\
$\quad$ $\bullet$ $10^{-2} < a_{1,f}/\mathrm{AU} < 12$ & \maintable{24}{0} & \maintable{24}{1} & \maintable{24}{2} & \maintable{24}{3} & \maintable{24}{4} & \maintable{24}{5} & \maintable{24}{6} & \maintable{24}{7} & \maintable{24}{8} & \maintable{24}{9} & \maintable{24}{10} & \maintable{24}{11} & \maintable{24}{12} & \maintable{24}{13} & --- & --- \\
$\quad$ $\bullet$ $a_{1,f}/\mathrm{AU} < 10^{-2}$ & \maintable{25}{0} & \maintable{25}{1} & \maintable{25}{2} & \maintable{25}{3} & \maintable{25}{4} & \maintable{25}{5} & \maintable{25}{6} & \maintable{25}{7} & \maintable{25}{8} & \maintable{25}{9} & \maintable{25}{10} & \maintable{25}{11} & \maintable{25}{12} & \maintable{25}{13} & --- & --- \\
\midrule
Triple destabilization & {\bf \maintable{26}{0}} & {\bf \maintable{26}{1}} & \maintable{26}{2} & \maintable{26}{3} & \maintable{26}{4} & \maintable{26}{5} & \maintable{26}{6} & \maintable{26}{7} & \maintable{26}{8} & \maintable{26}{9} & \maintable{26}{10} & \maintable{26}{11} & \maintable{26}{12} & \maintable{26}{13} & --- & --- \\
\midrule[0.08pt]
$\bullet$ On MS & \maintable{27}{0} & \maintable{27}{1} & \maintable{27}{2} & \maintable{27}{3} & --- & --- & --- & --- & --- & --- & --- & --- & \maintable{27}{12} & \maintable{27}{13} & --- & --- \\
$\bullet $ Post-MS & \maintable{28}{0} & \maintable{28}{1} & \maintable{28}{2} & \maintable{28}{3} & \maintable{28}{4} & \maintable{28}{5} & \maintable{28}{6} & \maintable{28}{7} & \maintable{28}{8} & \maintable{28}{9} & \maintable{28}{10} & \maintable{28}{11} & \maintable{28}{12} & \maintable{28}{13} & --- & --- \\
\midrule[0.08pt]
$\quad$ $\bullet$ AGB + MS & \maintable{29}{0} & \maintable{29}{1} & \maintable{29}{2} & \maintable{29}{3} & \maintable{29}{4} & \maintable{29}{5} & \maintable{29}{6} & \maintable{29}{7} & --- & --- & --- & --- & \maintable{29}{12} & \maintable{29}{13} & --- & --- \\
\midrule[0.08pt]
$\bullet$ Compact Object & \maintable{30}{0} & \maintable{30}{1} & \maintable{30}{2} & \maintable{30}{3} & \maintable{30}{4} & \maintable{30}{5} & \maintable{30}{6} & \maintable{30}{7} & \maintable{30}{8} & \maintable{30}{9} & \maintable{30}{10} & \maintable{30}{11} & \maintable{30}{12} & \maintable{30}{13} & --- & --- \\
\midrule[0.08pt]
$\quad$ $\bullet$ CO WD + MS & \maintable{31}{0} & \maintable{31}{1} & \maintable{31}{2} & \maintable{31}{3} & \maintable{31}{4} & \maintable{31}{5} & \maintable{31}{6} & \maintable{31}{7} & --- & --- & --- & --- & \maintable{31}{12} & \maintable{31}{13} & --- & --- \\
$\quad$ $\bullet$ CO WD + AGB & \maintable{32}{0} & \maintable{32}{1} & \maintable{32}{2} & \maintable{32}{3} & \maintable{32}{4} & \maintable{32}{5} & \maintable{32}{6} & \maintable{32}{7} & \maintable{32}{8} & \maintable{32}{9} & \maintable{32}{10} & \maintable{32}{11} & \maintable{32}{12} & \maintable{32}{13} & --- & --- \\
$\quad$ $\bullet$ CO + CO WD & \maintable{33}{0} & \maintable{33}{1} & \maintable{33}{2} & \maintable{33}{3} & \maintable{33}{4} & \maintable{33}{5} & \maintable{33}{6} & \maintable{33}{7} & \maintable{33}{8} & \maintable{33}{9} & \maintable{33}{10} & \maintable{33}{11} & \maintable{33}{12} & \maintable{33}{13} & --- & --- \\
\bottomrule
\end{tabular}
\caption{\small Probabilities $P$ (in per cent) of various evolutionary channels for both sampled triple populations. Any two objects in the first column refer to inner binary components. For the post-MS and compact object mergers the integer numbers in brackets denote the stellar type of the remnant according to the binary evolution algorithm that occurs most frequently.  We refer to Table \ref{table:stdes} for the meaning of the stellar types and for abbreviations used and to Fig. \ref{figfrac} for a complete overview of all occurring combinations of stellar types prior to an inner binary merger. In the no merger channels $a_{1,f}$ denotes $a_1$ at $t = t_H$. The quantities $f_\mathrm{KCTF}$, $f_\mathrm{CE}$ and $f_\mathrm{EC}$ denote the fraction of systems within each channel for which significant KCTF, CE evolution and a merger as a result of a highly eccentric collision applies respectively. For $f_\mathrm{KCTF}$ a distinction is made between the type of the star in which significant tidal energy is dissipated: MS (both stars are MS stars or the primary is a compact object and the secondary is a (low-mass) MS star), PHG/PRGB/PAGB (primary star is a HG/RGB/AGB star) and SHG/SRGB/SAGB (secondary star is a HG/RGB/AGB star). A long dash (---) indicates that this particular combination is not applicable. }
\label{table:outcomes}
\end{table*}

\begin{figure*}
\center
\includegraphics[scale = 0.51, trim = 0mm 0mm 0mm 0mm]{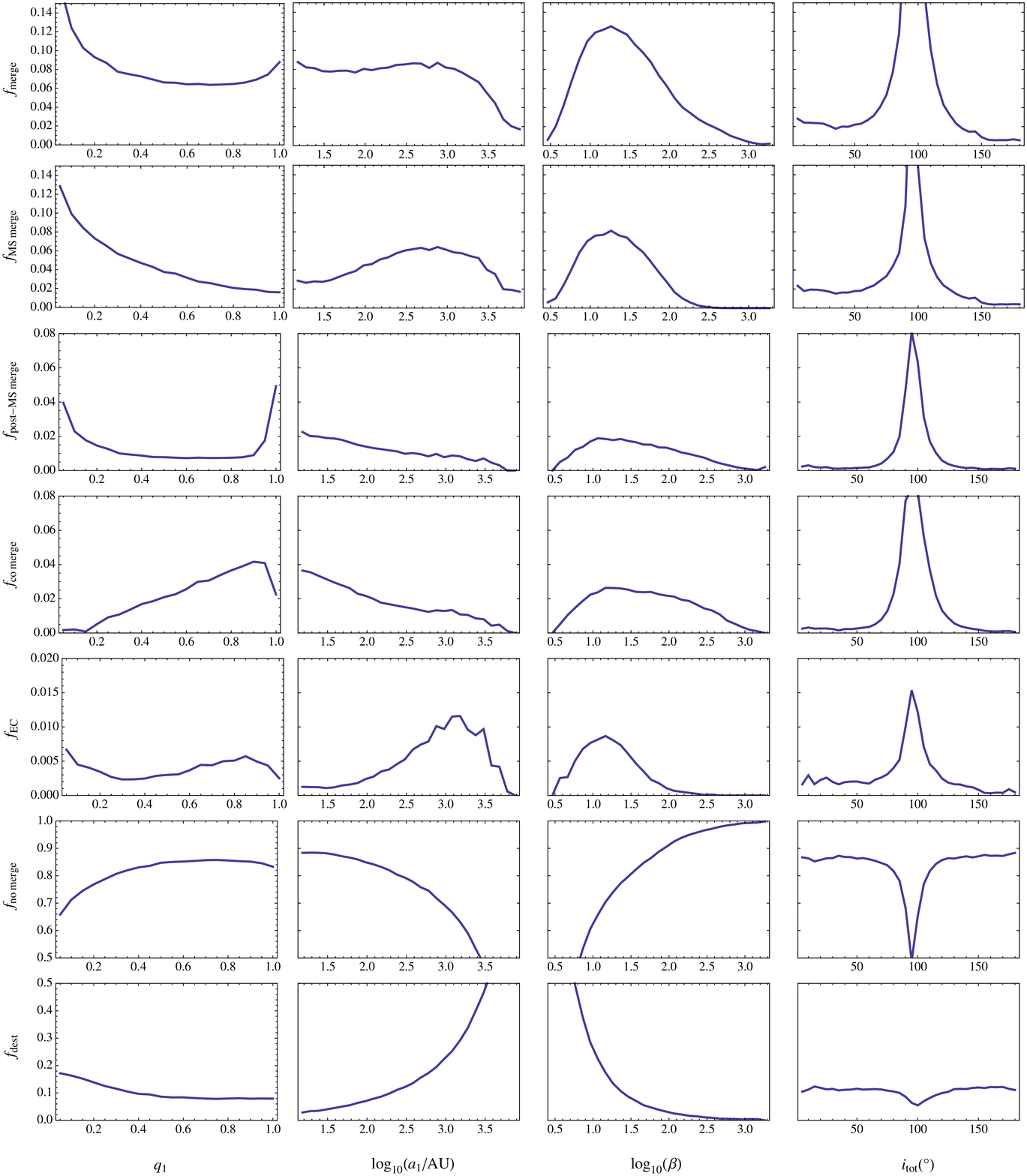}
\caption{\small Probabilities of the main channels, expressed as fractions of all systems, as a function of initial parameters for TSM1; $q_1 \equiv m_2/m_1$ and $\beta \equiv a_2/a_1$. Abbreviations used: co - compact object; dest - destabilization; EC - eccentric collision. In the last row only post-MS and compact object mergers are included. }
\label{fig:fracinitdep}
\end{figure*}

We present the main channels that we find in our population synthesis study in Table \ref{table:outcomes}, where the first two columns show the probabilities $P$ (in per cent) of the main channels and several sub channels of both sampled populations TSM1 and TSM2. In addition to this table, Fig. \ref{figfrac} shows more detailed information on the likelihood of the different combinations of inner binary mergers where the stellar types used are defined in Table \ref{table:stdes}. A post-MS merger is defined as a merger between a post-MS (non-compact object) primary and any secondary, whereas a compact object merger is defined as a merger between a compact object primary and any secondary. In addition to the probabilities of the channels information on the occurrence of KCTF, CE evolution and the nature of the merger (i.e. occurring in a circular or highly eccentric orbit) is given by means of the quantities $f_\mathrm{KCTF}$, $f_\mathrm{CE}$ and $f_\mathrm{EC}$. We define $f_\mathrm{KCTF}$ as the fraction of systems in each channel in which $|\dot{e}_{1,\mathrm{STD}}| > 10^{-18} \, \mathrm{s^{-1}}$ (such that there is a significant change in $e_1$ due to STD in a Hubble time) and $0.1 \, |\dot{e}_{1,\mathrm{TF}}| < |\dot{e}_{1,\mathrm{STD}}| < |\dot{e}_{1,\mathrm{TF}}|$ (i.e. $\dot{e}_{1,\mathrm{STD}}$ and $\dot{e}_{1,\mathrm{TF}}$ are of comparable order of magnitude). This can happen at various points in the evolution and is strongly influenced by the envelope structure of the inner binary stars because this determines the effectiveness with which tidal energy can be dissipated. In particular if any inner binary star possesses a convective envelope (most notably low-mass MS, HG, RGB and AGB stars) strong tidal friction is likely if $e_1$ is also high due to Kozai cycles. In Table \ref{table:outcomes} a distinction is made between the types of the star in which tidal energy is dissipated for the duration of KCTF. Furthermore, $f_\mathrm{CE}$ is defined as the fraction of systems in each channel in which CE evolution is at some point invoked. This includes both CE evolution triggered in highly eccentric orbits and CE evolution triggered by RLOF (in circular orbits). The fraction $f_\mathrm{CE}$ includes the possibility for multiple phases of CE evolution. Lastly, the quantity $f_\mathrm{EC}$ is defined as the fraction of systems in each channel in which a merger occurs due to an eccentric collision. We define such a collision to occur if the inner binary periastron distance is smaller than or equal to the sum of the inner binary radii, i.e. if $a_1(1-e_1) \leq R_1+R_2$ and when in the corresponding circular case there would not be a collision, i.e. $a_1 > R_1+R_2$. 

An important result of our simulations shown in Table \ref{table:outcomes} is that in $\approx 8 \%$ of all systems an inner binary merger occurs, in $\approx 6 \%$ of all systems no inner binary merger occurs but the orbit is shrunk significantly and that $\approx 10\%$ of all systems become dynamically unstable at some point in the evolution. This means that in $\approx 24 \%$ of all systems the tertiary significantly alters the inner binary evolution whereas in the absence of the tertiary there would not be such interaction. Taking into account that systems in which the inner binary components merge during the MS ($\approx 4\%$) or in which a dynamical instability occurs during the MS ($\approx 4 \%$) are not likely to be observed (cf. Sect. \ref{sect:results:mergers}) this estimate is reduced by $\approx 8\%$. This implies that the tertiary significantly alters the inner binary evolution in $\approx 16 \% /(1-0.08) \approx 17\%$ of the ``observable'' triples, i.e. triples in which the hierarchical structure is not disrupted in an early stage in the evolution.

A second important feature evident from Table \ref{table:outcomes} is that the results for TSM1 and TSM2 are generally similar, even though they have been sampled by different methods. To provide insight into the dependence of the probabilities of the main channels on the initial parameters we show in Fig. \ref{fig:fracinitdep} these probabilities (expressed as fractions) as function of the initial parameters $q_1 \equiv m_2/m_1$, $a_1$, $\beta$ and $i_\mathrm{tot}$. We will refer to this figure and Table \ref{table:outcomes} when discussing the three main channels in more detail below.

\subsection{Inner binary mergers}
\label{sect:results:mergers}
As a consequence of high-amplitude eccentricity cycles induced in the inner orbit the components in the inner binary may at some point merge, either directly because of an orbital collision (included in Table \ref{table:outcomes} in the fraction $f_\mathrm{EC}$) or indirectly as a result of strong tidal friction induced by high eccentricity, possibly invoking CE evolution. In the latter case the inner binary can either merge during the CE or survive the CE in a tight orbit and (possibly after another CE invoked by the secondary) subsequently merge as a result of orbital energy loss due to GW emission. We separate further discussion into mergers occurring on the MS, after the MS and mergers involving a primary compact object. 

\subsubsection{MS mergers}
\label{sect:results:mergers:MS}
As Table \ref{table:outcomes} demonstrates most mergers occur on the MS and for almost all of the latter the merger occurs through an eccentric collision. In the latter case $-4 \lesssim \log_{10}(1-e_1) \lesssim -5$ just prior to merger. In the {\tt Binary\_c} algorithm it is assumed that the result of the merger is a single MS star with mass $m_1 + m_2$; depending on the latter value the final remnant is either a CO WD (most cases), an ONe WD or a NS (following a core-collapse supernova). We find that octupole terms in the STD are very important for MS mergers, which is reflected by the large values of $|\epsilon_\mathrm{oct}|$ (cf. Eq. \ref{eq:eoct}) prior to merger for these systems; the latter quantity is narrowly distributed around $|\epsilon_\mathrm{oct}| \approx 10^{-1.5}$. This can be attributed to a combination of initially small semi-major axis ratios $\beta \equiv a_2/a_1$, small inner binary mass ratios $q_1 \equiv m_2/m_1$ and high outer orbit eccentricities $e_2$. Fig. \ref{fig:fracinitdep} indeed shows that the fraction of MS mergers, $f_\mathrm{MS \, merge}$, decreases significantly with increasing $q_1$ and is peaked towards small $\beta$ around $0.5 \lesssim \log_{10}(\beta) \lesssim 2.0$. The MS mergers are also characterized by high initial inclination angles (cf. Fig. \ref{fig:fracinitdep}). Note that a maximum in $f_\mathrm{MS \, merge}$ occurs at inclinations slightly above $i_\mathrm{tot} = 90^\circ$, i.e. for retrograde orbits, which is consistent with \citet{shapthomp12} who find an asymmetry in $\cos(i_\mathrm{tot})$ towards negative values for the probability of the eccentric Kozai mechanism to occur. With regards to the dependence of $f_\mathrm{MS \, merge}$ on $a_1$ it might be expected that MS mergers likely occur for small initial $a_1$ as a tighter orbit makes an eccentric collision more likely. However, we find that MS mergers on average have large initial $a_1$ with a maximum of $f_\mathrm{MS \, merge}$ around $a_1 \approx 10^{2.8} \, \mathrm{AU} \approx 6 \cdot 10^2 \, \mathrm{AU}$ (cf. Fig. \ref{fig:fracinitdep}). This implies that the effect of increasing the eccentricity maxima with decreasing $\beta$ (and thus increasing $a_1$) is stronger than the effect that eccentric mergers become more unlikely with increasing $a_1$ if $e_1$ were fixed. 

As Table \ref{table:outcomes} shows, for a small fraction of MS mergers (i.e. about $0.009$ of all MS mergers for TSM1) the merger does not occur due to high eccentricity, i.e. the merger occurs in a circular orbit. In these cases KCTF is responsible for strong orbital shrinkage leading to a tight inner orbit, eventually resulting in a merger in a circular orbit due to GW emission within a Hubble time. The fraction of MS merger systems in which this occurs (0.009 for TSM1) is indeed comparable with the fraction of MS merger systems that satisfy our KCTF criterion on the MS (0.010 for TSM1); this is also the case for TSM2. We find that such strong KCTF on the MS occurs only in low-mass inner binary systems ($m_1 < 1.25 \, M_\odot$) in which the components possess convective envelopes (i.e. the scenario of \citealt{fabrtr07}).

\begin{figure}
\center
\includegraphics[scale = 0.78, trim = 0mm 0mm 0mm 0mm]{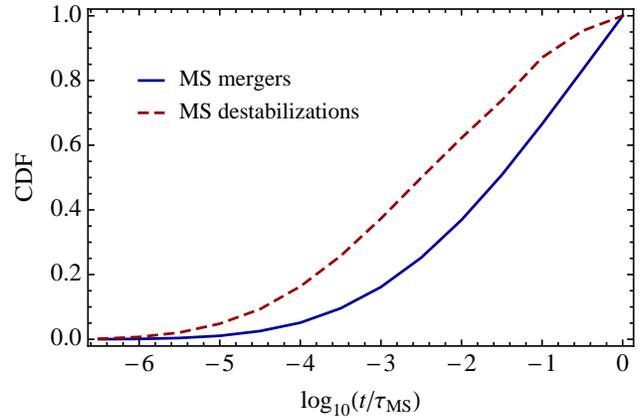}
\caption{\small Cumulative distributions for TSM1 of the times $t$ of MS merger (solid line) and MS destabilization (dashed line) normalized to the MS timescale, estimated as $\tau_\mathrm{MS} = 10 \, (m_1/M_\odot)^{-2.8} \, \mathrm{Gyr}$.}
\label{figMSdestime}
\end{figure}

The eccentric MS mergers, which constitute the bulk of MS mergers, typically occur early on the MS, with the majority of mergers (about 70\%) occurring within $10\%$ of the primary MS lifetime (Fig. \ref{figMSdestime}, solid line). This is because in many cases the collision occurs during the first few Kozai cycles which typically have timescales of a few Myr, a small fraction of the MS lifetime of the primary stars considered in the triple sample. The time of merger is thus mainly determined by the Kozai period $P_K \propto (P_2/P_1) \, P_2$ and hence the orbital periods. We indeed find that the distribution of the times of MS mergers is very similar to the initial distribution of the Kozai periods. The fact that most MS mergers occur early on the MS makes it unlikely to observe these systems as triples. A similar conclusion applies to the MS destabilizations (Sect. \ref{sect:results:destab}) that happen even earlier. 

A relevant question for the MS mergers is whether the merged MS remnant could evolve to a blue straggler star, i.e. a star more luminous than stars at the MS turn-off point. This triple blue straggler scenario has been proposed by \citet{per09}. If the MS stars merge very early in the evolution then the age of the merged MS star would appear to be consistent with its mass and so it would not be observed as a blue straggler star. For this reason most of the MS mergers in our sample are unlikely to be blue straggler progenitors.

\subsubsection{Post-MS mergers}
\label{sect:results:mergers:postMS}
For any merger after the MS an evolutionary change in the inner binary system plays a key role in the mechanism that leads to the merger. We progressively discuss the various post-MS merger scenarios of Table \ref{table:outcomes}. The radius expansion during the primary HG is responsible for an eccentric collision for most mergers during this phase. Note that in such an event the {\tt Binary\_c} algorithm invokes CE evolution, in contrast to the eccentric MS mergers. For about $20\%$ of the HG + MS mergers strong KCTF triggered by the substantial increase of the tidal strength quantity $k_{\mathrm{am},1}/T_1$ during the primary HG phase leads to orbital circularization and shrinkage to $a_1 \approx 0.4 \, \mathrm{AU}$. Subsequently, CE evolution is invoked, leaving a single HG star. For the RGB + MS mergers one might expect eccentric collisions to be important because the RGB phase is associated with a substantial increase in radius. However, the large increase in the tidal strength quantity $k_{\mathrm{am},1}/T_1$ during this phase is responsible for strong KCTF which circularizes and shrinks the inner orbit to $a_1 \approx 0.3 \, \mathrm{AU}$ leading to CE evolution, which happens in nearly all cases.  

During the CHeB phase (radiative envelope) tides are unimportant and most mergers are driven by radius expansion in conjunction with eccentricity cycles, in all cases leading to an eccentric merger. As expected, KCTF is once again important for primary AGB mergers (convective envelope). For these systems the secondary at the moment of merger is most likely a MS or a CHeB star. For primary AGB mergers KCTF is effective mainly during the primary AGB phase, although in about $25\%$ of these systems KCTF is also important during the primary RGB phase. After KCTF a CE is invoked; because the orbit at this point is tight the inner binary does not survive the CE phase and merges into an AGB star (in case of AGB + MS merger) or CHeB star (in case of AGB + CHeB merger). 

The latter AGB + CHeB mergers, as well as RGB + HG and RGB + RGB mergers, have initial distributions of $q_1$ peaked towards high values ($q_1 \gtrsim 0.9$), explaining the sharp increase of $f_\mathrm{post-MS \, merge}$ for $0.9 \lesssim q_1 \leq 1$ (Fig. \ref{fig:fracinitdep}). In addition, Fig. \ref{fig:fracinitdep} shows that post-MS mergers typically have small $\beta$ similarly to MS mergers, although $f_\mathrm{post-MS \, merge}$ shows a clear tail for larger $\beta$ which is not the case for $f_\mathrm{MS \, merge}$; for compact object mergers this tail is even more prominent. Furthermore, post-MS mergers tend to have smaller initial $a_1$ than the MS mergers. This can be understood by noting that KCTF is typically more effective for tighter inner binary systems.

\begin{figure}
\center
\includegraphics[scale = 0.78, trim = 0mm 0mm 0mm 0mm]{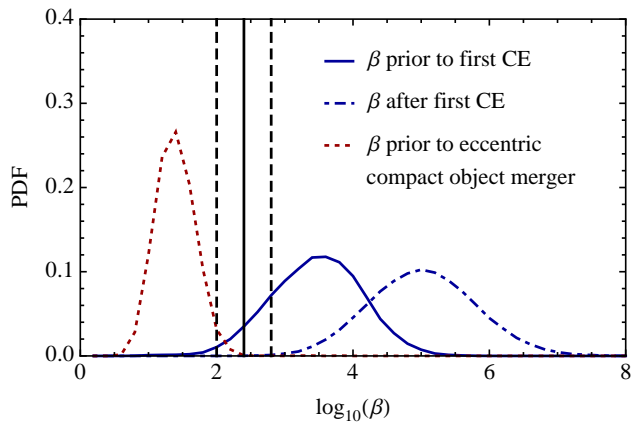}
\caption{\small Distribution for TSM1 of $\beta \equiv a_2/a_1$ prior (solid line) and after (dot-dashed line) the first CE for all systems in which CE evolution occurs and that survive the CE phase. Shown with vertical lines are the median values (solid) and spread values (dashed) of $\beta_\mathrm{crit,GR}$ (Eq. \ref{eq:betacritGR}) for the systems in which CE occurs, just after the CE (96\% of all cases fall within the indicated spread). Lastly, the dotted line shows the distribution of $\beta$ prior to merger for all eccentric compact object mergers. }
\label{fig:betaCE}
\end{figure}

\subsubsection{Compact object mergers}
\label{sect:results:mergers:compobj}
If KCTF triggers a CE during the inner binary giant phases and the inner binary system survives this CE, a possible outcome is a compact object merger with the primary either a He WD (KCTF during RGB) or a CO WD (KCTF during RGB/AGB). Fig. \ref{fig:betaCE} shows the distribution of $\beta$ just prior to and after the first CE for all systems in which CE evolution occurs and that survive the CE (this includes systems in which an inner binary merger does not occur). After the first CE $\beta \sim 10^5$, which is much larger than the critical value $\beta_\mathrm{crit,GR}$ (cf. Eq. \ref{eq:betacritGR}) for which general relativistic precession dominates in these systems (Fig. \ref{fig:betaCE}, vertical lines). We therefore find that after the first CE $\beta$ is large enough that subsequent Kozai cycles do not occur. In other words, the inner binary evolution after the first CE is not affected by the tertiary. 

This subsequent evolution is driven by stellar evolution of the secondary (possibly invoking an additional CE) and/or GW emission if the inner orbit is sufficiently tight. The type of merger is mainly determined by the strength of KCTF: the stronger KCTF, the tighter the inner orbit after the first CE and the smaller the probability for the secondary to become a highly evolved star. This results in a multitude of compact object mergers, among which are He WD + HG/RGB mergers and CO WD + He WD mergers. The former are believed to be important channels leading to carbon-rich K-type giants, known as early-type R stars \citep{izz07}; the latter have been discussed as possible progenitors of H-deficient carbon-rich supergiants known as RCrB stars \citep{tiss09}.

\begin{figure}
\center
\includegraphics[scale = 0.78, trim = 0mm 0mm 0mm 0mm]{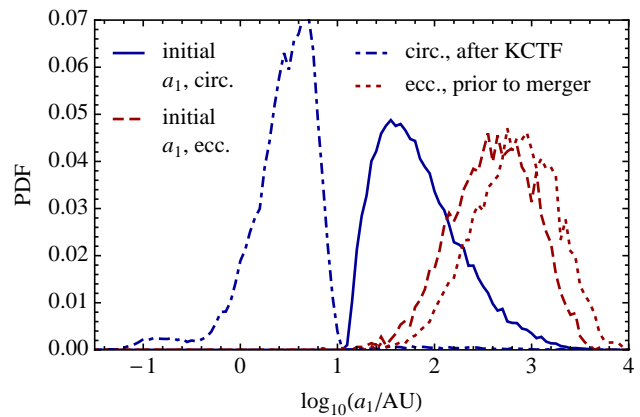}
\caption{\small Distribution for TSM1 of the initial $a_1$ for compact object mergers in which strong KCTF triggers a CE and finally a merger in a circular orbit (``circ.''; solid line) and in which an eccentric collision occurs (``ecc.''; dashed line). In addition, for the former systems the distribution is shown of $a_1$ after KCTF during the primary AGB phase (dot-dashed line); for the latter systems the distribution is shown of $a_1$ just prior to the eccentric collision (dotted line).  Note that the systems having $a_1 \sim 10^{-1} \, \mathrm{AU}$ after primary AGB KCTF are those in which strong KCTF occurred also during an earlier stage, most notably the primary RGB phase. }
\label{fig:a1distcompTICMTIEM}
\end{figure}

For a fraction $f_\mathrm{EC} \approx 0.1$ of all compact object mergers strong KCTF and CE evolution during the MS and post-MS phases are avoided. These systems are typically initially wide ($a_1 \gtrsim 10^2 \, \mathrm{AU}$; cf. Fig. \ref{fig:a1distcompTICMTIEM}, dashed line) and expand to even wider orbits during the AGB phases due to wind mass loss (Fig. \ref{fig:a1distcompTICMTIEM}, dotted line), as opposed to shrinking as a consequence of KCTF. Similar to the example system in Sect. \ref{sect:extrev:TIEM} the ratio $\beta \equiv a_2/a_1$ decreases because of this mass loss. Fig. \ref{fig:betaCE} (dotted line) shows that the resulting $\beta$ is narrowly distributed around $\beta \approx 15$. Octupole order effects are very important in such systems, with $|\epsilon_\mathrm{oct}|$ peaking around $\approx 10^{-2}$. In addition, because at this stage the inner orbit semi-major axis is large (typically $a_1 \approx 10^3 \, \mathrm{AU}$; cf. Fig. \ref{fig:a1distcompTICMTIEM}, dotted line) there is no significant damping of Kozai cycles due to additional sources of inner orbit precession. Consequently, in such systems orbital flips can occur as in Sect. \ref{sect:extrev:TIEM} during which extremely high eccentricities can be reached (e.g. \citealt{naoz11,lithnaoz11, shapthomp12}), in our calculations as high as $1-e_1 \sim 10^{-8}$. 

In most cases these high eccentricities lead to orbital collisions. The value of the eccentricity required for such a collision is determined by $a_1$ and the radius and hence the type/mass of the secondary object (mainly a MS star, a CHeB star or a CO WD) and varies between $1-e_1 \sim 10^{-3}$ (CHeB companion), $1-e_1 \sim 10^{-5}$ (MS companion) and $1-e_1 \sim 10^{-7}$ (CO WD companion). We will discuss the CO WD + CO WD eccentric merger scenario in more detail in Sect. \ref{sect:implications}. 

In a few cases, mainly for CO WD + low-mass MS systems, the high eccentricities do not lead to eccentric collisions but induce very strong tidal friction in the low-mass secondary that possesses a convective envelope, thus circularizing and shrinking the inner orbit to $a_1 \approx 10^{-2} \, \mathrm{AU}$. These cases correspond to the MS $f_\mathrm{KCTF}$ fractions for the CO WD + MS merger channel and the no merger channel with $a_{1,f} < 12 \, \mathrm{AU}$ in Table \ref{table:outcomes}. Depending on the precise values of $a_1$ and the masses the subsequent tight binary does or does not merge within a Hubble time due to GW emission. The systems in which the latter scenario applies constitute about $20\%$ of the tightest no inner binary merger systems ($a_{1,f} < 10^{-2} \, \mathrm{AU}$ in Table \ref{table:outcomes}). They may be recognized in Table \ref{table:outcomes} as the no merger systems with $a_{1,f} < 10^{-2} \, \mathrm{AU}$ in which a CE does not occur.

\subsection{No inner binary mergers}
\label{sect:results:nomergers}

\begin{figure}
\center
\includegraphics[scale = 0.78, trim = 0mm 0mm 0mm 0mm]{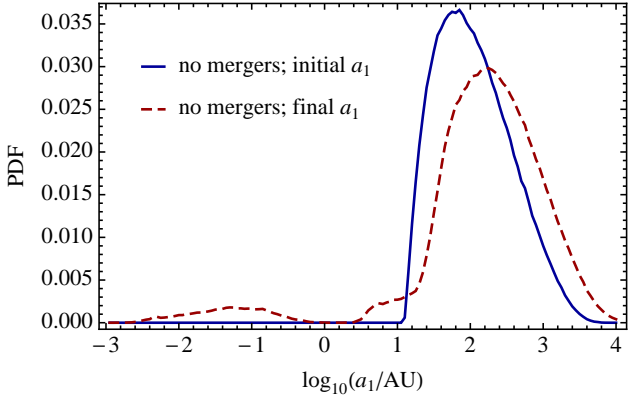}
\caption{\small Distribution for TSM1 of the initial (solid line) and final $a_1$ (dashed line) for all no merger systems. }
\label{fignomergea1}
\end{figure}

If the initial inclination angle is not close to $90^\circ$ and/or the initial $\beta$ is large then the maximum eccentricity induced by Kozai cycles during the evolution may not be high enough to drive an inner binary merger. This is demonstrated by Fig. \ref{fig:fracinitdep} which shows that $f_\mathrm{no \, merge}$ increases strongly with increasing $\beta$ ($f_\mathrm{no \, merge} \rightarrow 1$ for sufficiently large $\beta$) and that systems with no or moderate inclination are likely not to merge. Furthermore, $f_\mathrm{no \, merge}$ increases with decreasing $a_1$ which might be somewhat counterintuitive. This can be understood by considering that as $a_1$ decreases $\beta$ is also generally larger, thus decreasing the eccentricity maxima.

Although by definition a merger does not occur for these systems, it is possible that the inner orbit is shrunk significantly due to KCTF as is illustrated by Table \ref{table:outcomes}: in about $6 \%$ of all systems the final inner orbit semi-major axis is smaller than 12 AU, which is the minimum possible value in the absence of the tertiary (cf. Sect. \ref{sect:popsyn:selcrit}). In the majority of these systems the main points in the evolution at which KCTF is important are the primary RGB and AGB phases occurring in roughly equal proportion. Fig. \ref{fignomergea1} compares the initial and final distributions of $a_1$. The final distribution shows a broad peak mainly due to strong KCTF around $a_{1,f} \sim 10^{-1.5} \, \mathrm{AU} \approx 0.03 \, \mathrm{AU}$ ($P_1 \sim 2 \, \mathrm{d}$). For the tightest of these systems ($a_{1,f} < 10^{-2} \, \mathrm{AU}$) it is also possible that very high eccentricity cycles lead to a tidal capture if the systems consists of a CO WD + low-mass MS star, which is the same scenario discussed in the last paragraph of Sect. \ref{sect:results:mergers:compobj}. 

\begin{figure}
\center
\includegraphics[scale = 0.78, trim = 0mm 0mm 0mm 0mm]{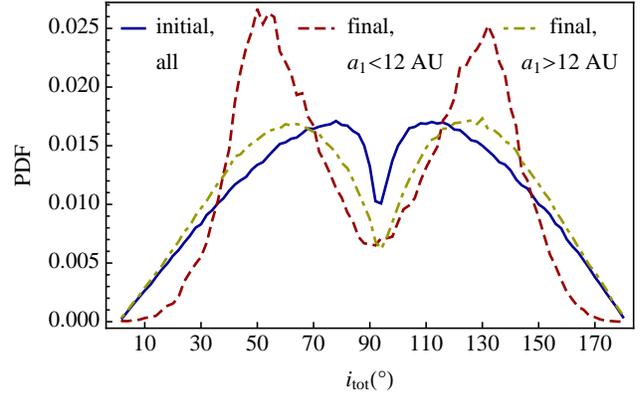}
\caption{\small Distribution for TSM1 of the initial inclination angle for the no merger systems (solid line) and the final inclination angle for the no merger systems with final $a_1 < 12 \, \mathrm{AU}$ (dashed line) and $a_1 > 12 \, \mathrm{AU}$ (dot-dashed line). }
\label{fignomergeitot}
\end{figure}

\pgfplotstableread{tablenomerge.txt}{\filetablenomerge}
\newcommand{\tablenomerge}[2]{\pgfplotstablegetelem{#1}{#2}\of{\filetablenomerge}  \pgfplotsretval}

\begin{table}
\begin{center}
\begin{tabular}{lcccccc}
\toprule 
& \multicolumn{2}{c}{Wide} & \multicolumn{2}{c}{Tight} & \multicolumn{2}{c}{Very tight} \\
\cmidrule(r){2-3} \cmidrule(r) {4-5} \cmidrule(r){6-7}
& TSM1 & TSM2 & TSM1 & TSM2 & TSM1 & TSM2 \\
\midrule
CO + CO WD & \tablenomerge{0}{0} & \tablenomerge{0}{1} & \tablenomerge{0}{2} & \tablenomerge{0}{3} & \tablenomerge{0}{4} & \tablenomerge{0}{5} \\  
CO WD + MS & \tablenomerge{1}{0} & \tablenomerge{1}{1} & \tablenomerge{1}{2} & \tablenomerge{1}{3} & \tablenomerge{1}{4} & \tablenomerge{1}{5} \\   
CO + He WD & \tablenomerge{2}{0} & \tablenomerge{2}{1} & \tablenomerge{2}{2} & \tablenomerge{2}{3} & \tablenomerge{2}{4} & \tablenomerge{2}{5} \\   
He + He WD & \tablenomerge{3}{0} & \tablenomerge{3}{1} & \tablenomerge{3}{2} & \tablenomerge{3}{3} & \tablenomerge{3}{4} & \tablenomerge{3}{5} \\   
He WD + MS & \tablenomerge{4}{0} & \tablenomerge{4}{1} & \tablenomerge{4}{2} & \tablenomerge{4}{3} & \tablenomerge{4}{4} & \tablenomerge{4}{5} \\   
\bottomrule
\end{tabular}
\end{center}
\caption{\small Likelihood of the main inner binary configurations at the end of the evolution for the no merger systems. These systems are divided into three groups, based on the final value of $a_1$: wide ($a_{1,f} > 12 \, \mathrm{AU}$), tight ($10^{-2} <  a_{1,f}/\mathrm{AU} < 12$) and very tight ($a_{1,f} < 10^{-2} \, \mathrm{AU}$). The fractions are relative to the total number of systems in each group. }
\label{table:outcomesnomerge}
\end{table}

The final distribution of $a_1$ shows a gap around $a_{1,f} \sim 1 \, \mathrm{AU}$. This gap separates systems in which KCTF is followed by a CE ($a_{1,f} < 1 \, \mathrm{AU}$) and in which a CE does not occur after KCTF ($a_{1,f} > 1 \, \mathrm{AU}$). In the latter case KCTF is responsible for an enhancement of systems near final values of $a_1 \sim 10 \, \mathrm{AU}$. If a CE occurs then this affects the final configuration of the inner binary system. This is shown in Table \ref{table:outcomesnomerge} where the likelihoods of the most important inner binary configurations at the end of the evolution are shown for the no merger systems. A distinction is made between the three groups with $a_{1,f} > 12 \, \mathrm{AU}$ (``wide''), $10^{-2} <  a_{1,f}/\mathrm{AU} < 12$ (``tight'') and $a_{1,f} < 10^{-2} \, \mathrm{AU}$ (``very tight''). A CE does not occur in the wide systems and therefore these end mainly as CO WD + MS and CO WD + CO WD systems. In the case of a CE there is also a possibility for forming a He WD. The tighter systems can therefore also end as CO WD + He WD, He WD + He WD and He WD + MS systems. 

The no merger systems are interesting observationally because the hierarchical structure remains intact during the entire evolution and the process of KCTF distinctly affects the orbital inclination angle. Typically during episodes of KCTF $e_1$ gradually approaches its maximum value in the Kozai cycle, followed by rapid circularization. The inclination angle during this process tends to remain fixed at its value associated with the eccentricity maximum, i.e. close to one of the two critical values determined by $\cos^2(i_\mathrm{tot}) = 3/5$. Therefore an observational marker of systems in which KCTF has once been important is an inclination angle close to these critical values (see also \citealt{fabrtr07}). Fig. \ref{fignomergeitot} shows the initial and final inclination angles for the no merger systems, distinguishing between $a_{1,f} > 12 \, \mathrm{AU}$ and $a_{1,f} < 12 \, \mathrm{AU}$. The initial distribution shows a clear lack of systems near $i_\mathrm{tot} \approx 90^\circ$ as also illustrated by Fig. \ref{fig:fracinitdep}. Systems for which strong KCTF applies at some point in the evolution ($a_{1,f} < 12 \, \mathrm{AU}$) end with inclination angles strongly peaked towards the critical values, whereas for systems in which KCTF is weaker or does not act at all (final $a_1 > 12 \, \mathrm{AU}$) these peaks are much less prominent. We therefore expect triples with tight inner orbits to show a different inclination distribution than those with much wider inner orbits. 

\subsection{Triple destabilizations}
\label{sect:results:destab}
If the initial $\beta$ is particularly low, typically $\beta \lesssim 10$ (cf. Fig. \ref{fig:fracinitdep}), then the triple system may at some point in the evolution become dynamically unstable. Subsequent evolution is not modeled by our triple evolution algorithm because the secular evolution equations no longer apply in this case and three-body simulations are needed for an accurate description of the dynamical evolution. This scenario has been studied in detail by \citet{perkr12}, who refer to it as the triple evolution dynamical instability. In our simulations the destabilization occurs mainly during the MS or when the primary is an AGB star or a CO WD (cf. Table. \ref{table:outcomes}). 

In the case of MS destabilization the triple system is initially marginally stable (i.e. only just satisfies $\beta > \beta_\mathrm{crit}$, cf. Sect. \ref{sect:triplealg:binalg}) but due to octupole-order terms of the STD, which are important since $\beta$ is (very) small and/or $e_2$ is high, $e_2$ varies periodically until it reaches a value high enough such that $\beta \leq \beta_\mathrm{crit}$, i.e. a triple destabilization. The time when this occurs is determined by the Kozai period $P_K$. Similarly to the MS mergers, this occurs early in the evolution with most ($90\%$) destabilizations occurring within $10\%$ of the primary MS lifetime (cf. Fig. \ref{figMSdestime}). 

In the other cases destabilization is triggered by mass loss in the inner orbit which, if fast and isotropic, acts to decrease $\beta$ (i.e. the same mechanism discussed in the context of eccentric compact object mergers in Sect. \ref{sect:results:mergers}) to a point where $\beta \leq \beta_\mathrm{crit}$. This happens when the primary loses a significant amount of mass as it evolves from the AGB phase to a WD and similarly when this happens to the secondary. In a small number of cases both inner binary components are CO WDs when the instability occurs and since there exists a finite probability of collision in the triple evolution dynamical instability (approximately 0.1 as found by \citealt{perkr12}) this could potentially lead to a CO WD collision. This scenario is included in  Sect. \ref{sect:implications}.

\section{Implications of triple-induced CO WD mergers}
\label{sect:implications}
As mentioned in the introduction the CO WD merger channel is considered as an important candidate progenitor channel for SNe Ia (see \citealt{maozman12} and \citealt{wanghan12} for recent reviews). In this section we explore the implications of triple-induced CO WD mergers by estimating the expected SN Ia rate. First we briefly discuss the triple-induced CO WD merger scenarios that we find in our population synthesis study (Sect. \ref{sect:implications:scenarios}). In order to estimate the SNe Ia rate we must make assumptions for which combinations of CO WD total masses and mass ratios prior to merger lead to a SN Ia explosion. We base these assumptions on the results of various detailed simulations of CO WD mergers (Sect. \ref{sect:implications:likelihood}). With these assumptions and by normalizing the results we obtain the expected SNe Ia rates. We then compare these rates to the binary population synthesis study of Claeys et al. (in prep.) and to those inferred from observations (Sect. \ref{sect:implications:rates}). 

\subsection{Scenarios}
\label{sect:implications:scenarios}
In Sect. \ref{sect:results} we have found that in the sampled triple populations CO WD mergers can be the result of 1) strong KCTF followed by one or multiple CE phases eventually followed by a merger in a circular orbit due to GW emission, 2) extremely high eccentricities ($1-e_1 \sim 10^{-7}$) excited in wide inner binaries with a distant companion where KCTF is avoided, resulting in an orbital collision and 3) triple destabilization driven by mass loss during the inner binary AGB phases, possibly leading to a collision \citep{perkr12}. Note that systems in the first scenario typically have initial $a_1$ in the range $10^1 \lesssim a_1 /\mathrm{AU} \lesssim 10^2$ whereas for the second scenario this range is $10^2 \lesssim a_1 /\mathrm{AU} \lesssim 10^3$ (cf. Fig. \ref{fig:a1distcompTICMTIEM}). As shown in Table \ref{table:outcomes} scenario 1 is about 10 times more likely than scenario 2 (i.e. $f_\mathrm{EC} \approx 0.1$ for the CO WD merger channel), whereas scenario 3, taking into account that a triple instability leads to a collision in about 0.1 of all cases \citep{perkr12}, is about 10 times less likely than scenarios 1 and 2 combined. Note that in the absence of the tertiary these channels would not exist for the sampled populations, i.e. they are all induced by the secular gravitational influence of the tertiary. In binary population synthesis studies such channels are therefore not taken into account. 

\begin{figure}
\center
\includegraphics[scale = 0.75, trim = 0mm 0mm 0mm 0mm]{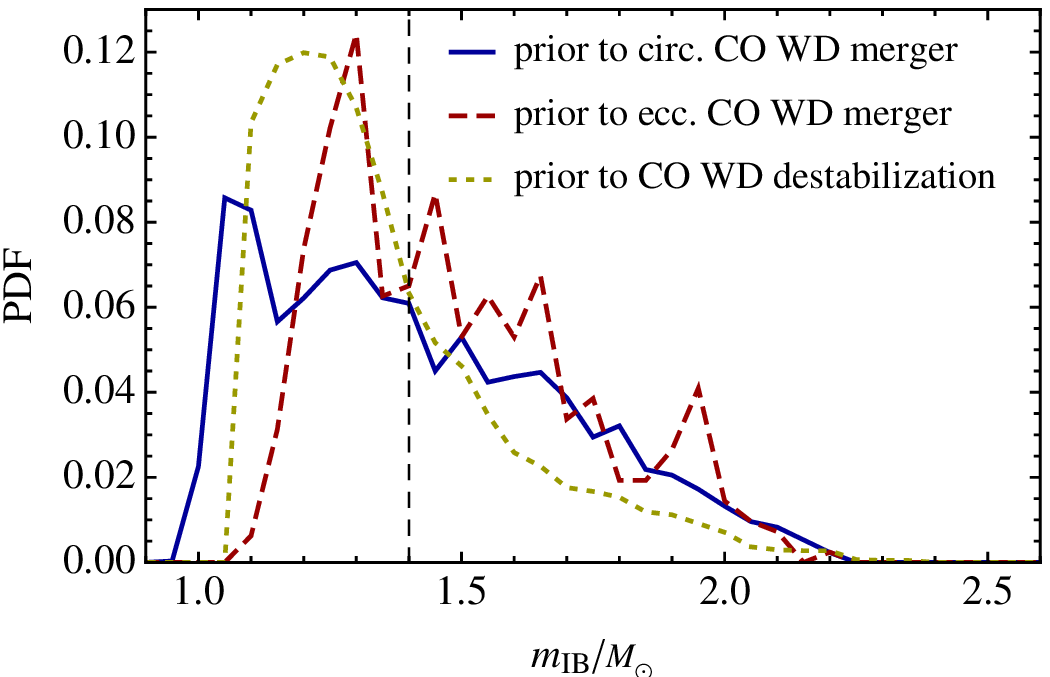}
\includegraphics[scale = 0.77, trim = 0mm 0mm 0mm 0mm]{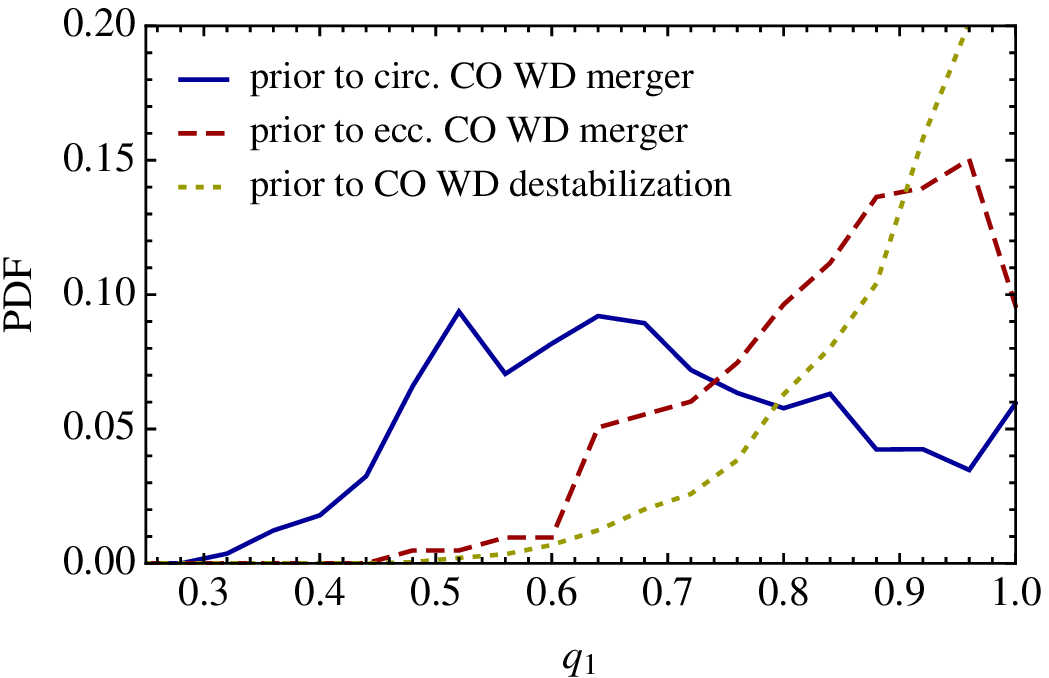}
\caption{\small Distribution for TSM1 of the inner binary total mass $m_\mathrm{IB} \equiv m_1+m_2$ (top; vertical dashed line denotes $M_\mathrm{Ch}$) and mass ratio $q_1 = m_2/m_1$ (bottom) just prior to CO WD merger or triple destabilization, for the CO WD inner binary circular (solid line) and eccentric (dashed line) mergers and the CO WD destabilizations (dotted line). }
\label{figCOCOmtotq}
\end{figure}

\subsection{Likelihood of a SN Ia explosion}
\label{sect:implications:likelihood}
Many uncertainties exist as to which configurations of CO WD mergers lead to a SN Ia explosion. Nevertheless, estimates of the minimum inner binary total mass $m_\mathrm{IB} \equiv m_1+m_2$ and mass ratio $q_1 \equiv m_2/m_1$ required can be inferred from detailed simulations. Here it is important to distinguish between mergers occurring in circular orbits (with GW emission driving the merger) and mergers occurring in highly eccentric orbits, very similar to head-on collisions. The latter are more likely to result in a SN Ia explosion because of the high relative speed, on the order of the escape speed ($\sim 10^3 \, \mathrm{km/s}$), which causes shocks that aid the ignition of the thermonuclear explosion. For circular mergers the minimum $q_1$ could be approximately $q_1 \approx 0.8$ (e.g. \citealt{pak11}) provided that $m_\mathrm{IB} > M_\mathrm{Ch} \approx 1.4 \, M_\odot$. Furthermore, for eccentric mergers a minimum mass ratio of $q_1 = 2/3$ can be inferred from Table 1 of \citet{ross09} and $q_1 = 0.6$ from Table 2 of \citet{raskin10}. According to the latter study the minimum total mass is between 1.0 and 1.2 $M_\odot$.

Fig \ref{figCOCOmtotq} shows the distributions of $m_\mathrm{IB}$ (top) and $q_1$ (bottom) prior to merger for the three triple-induced CO WD merger scenarios. In the circular merger scenario the merger can be both sub-$M_\mathrm{Ch}$ (57\%) and super-$M_\mathrm{Ch}$ (43\%) and $q_1$ is broadly distributed with $0.4 \lesssim q_1 \lesssim 1.0$. In the eccentric merger scenario $m_\mathrm{IB}$ exceeds the upper limit of the minimum mass for a SN Ia event of $1.2 \, M_\odot$ according to \citet{raskin10} for most systems (95\%), whereas $q_1$ is peaked towards high values, $0.6 \lesssim q_1 \lesssim 1$. Lastly, in the destabilization scenario $m_\mathrm{IB}$ is typically slighter lower than in the eccentric merger scenario whereas $q_1$ is even more strongly peaked towards high values.

The fact that $q_1$ in scenario 1 can be moderately small suggests that not all mergers in this scenario lead to a SN Ia explosion. In binary population synthesis studies a distinction in mass ratio is not usually made, however (see \citealt{chen12} for a study of the effect on the binary population synthesis rates of including more detailed constraints). In order to facilitate a comparison with results from such binary population synthesis studies we shall therefore disregard the mass ratio criterion and only consider a restriction on $m_\mathrm{IB}$: in Sect. \ref{sect:implications:rates} we shall assume that  a SN Ia explosion is the result in scenario 1 (circular merger) if $m_\mathrm{IB} > M_\mathrm{Ch}$ and in scenarios 2 and 3 (collision) if $m_\mathrm{IB} > 1.2 \, M_\odot$. 

\subsection{Expected SNe Ia rates}
\label{sect:implications:rates}
\begin{figure}
\center
\includegraphics[scale = 0.78, trim = 0mm 0mm 0mm 0mm]{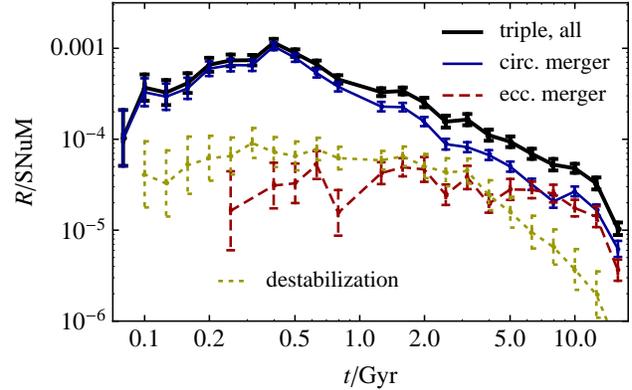}
\caption{\small SNe Ia delay time distribution (DTD) for TSM1 of the triple-induced inner binary CO WD merger and destabilization channels with the criteria for producing a SNe Ia event discussed in Sect. \ref{sect:implications:likelihood}. Shown are the combined rate (heavy solid line), the rate due to inner binary circular mergers with $m_\mathrm{IB} > M_\mathrm{Ch}$ (thin solid line) and eccentric mergers with $m_\mathrm{IB} > 1.2 \, M_\odot$ (dashed line) and the rate due to CO WD collisions induced by triple destabilizations assuming that in $10\%$ of all inner binary CO WD destabilizations a collision subsequently occurs (dotted line).}
\label{figDTD}
\end{figure}

Conforming to the procedure commonly employed in binary population synthesis studies, we normalize the number of systems in which a SN Ia event is expected to occur to the part of the galactic mass represented by the sampled systems. The details of this normalization are included in Appendix \ref{app:DTDcalc}; here we mention our most important assumptions. Firstly, it must be taken into account that the sampled triple populations constitute only a small fraction of all triple systems (cf. Sect. \ref{sect:popsyn:selcrit}). In addition we must assume specific values for the binary and triple fractions of the galactic population. The binary fraction $\alpha_\mathrm{bin}$ is a function of mass and is found to range between $\alpha_\mathrm{bin} = 0.56$ for solar-like stars \citep{rag10} to $\alpha_\mathrm{bin} \geq 0.7$ for O \& B stars \citep{kouw07}. For simplicity we assume here an intermediate constant value of $\alpha_\mathrm{bin} = 0.60$. Statistics of the triple fraction $\alpha_\mathrm{tr}$, which is similarly a strong function of mass, are even less well-constrained. The observed triple fractions range from 0.11 for solar-like stars \citep{rag10,ras10} to 0.5 for more massive B stars \citep{remev11}. Again neglecting the mass dependence we take a conservative intermediate constant value $\alpha_\mathrm{tr} = 0.25$, where $\alpha_\mathrm{tr}$ is not to be understood as a subset of $\alpha_\mathrm{bin}$. We neglect higher-order multiplicities, thus giving a single star fraction of $1 - \alpha_\mathrm{bin} - \alpha_\mathrm{tr} = 0.15$. With these fractions we calculate the total mass represented by the number of triple systems we have calculated and with which we normalize the SNe Ia rates for TSM1 (Eq. \ref{eq:DTDnorm:totmass}). We express the result in rate per $10^{10} \, M_\odot$ per century (SNuM). Fig. \ref{figDTD} shows these rates as a function of time $t$ after an assumed starburst at $t=0$, i.e. the delay time distribution (DTD). The contributions of the three scenarios of Sect. \ref{sect:implications:scenarios} taking into account the assumptions of Sect. \ref{sect:implications:likelihood} are indicated separately. 

Fig. \ref{figDTD} shows that the main contribution to the triple-induced rate is from the inner binary circular merger channel. The inner binary eccentric merger channel contributes significantly at late times ($t \sim 10 \, \mathrm{Gyr}$), whereas it does not at earlier times. This is because after the CO WD formation the Kozai period $P_K$ is typically quite long (the distribution of $P_K$ peaks around $P_K \approx 0.3 \, \mathrm{Gyr}$; see also Sect. \ref{sect:extrev:TIEM}) and it typically takes at least a few cycles before an orbital flip occurs (cf. Fig. \ref{fig:ex42406}, bottom right panel). Such an orbital flip plays a crucial role in driving eccentricities high enough for a collision in these systems ($1-e_1 \lesssim 10^{-7}$). The destabilization channel does not contribute significantly to the total rate at any time. 

\begin{figure}
\center
\includegraphics[scale = 0.78, trim = 0mm 0mm 0mm 0mm]{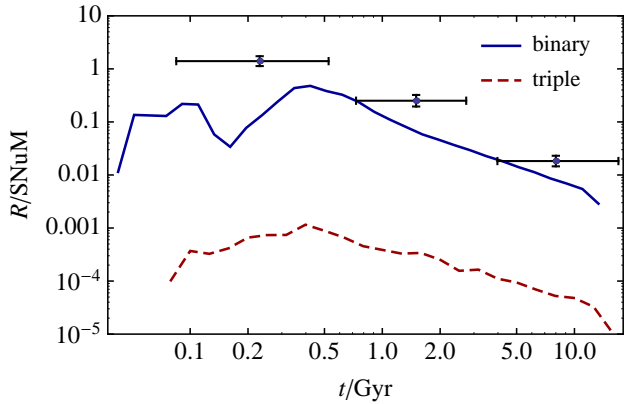}
\caption{\small SNe Ia DTD according to all triple-induced channels of Fig. \ref{figDTD} (dotted line) and according to the binary population synthesis study of Claeys et al. (in prep.) (solid line). In addition, observed SNe Ia rates (points with error bars) are shown from \citet{maozmanbrandt12}. }
\label{figDTDcomp}
\end{figure}

Furthermore, we show in Fig. \ref{figDTDcomp} the combined expected triple-induced SNe Ia rate according to the three channels discussed above (dashed line), the predicted SNe Ia DTD from the binary population synthesis study of Claeys et al. (in prep.) (solid line) and the observed rates \citep{maozmanbrandt12}. Note that in Claeys et al. (in prep.) a binary fraction of $\alpha_\mathrm{bin} = 1$ is assumed and that contributions are shown from both the single degenerate channel (single CO WD accreting material from a non-degenerate donor star until it reaches $M_\mathrm{Ch}$ and explodes; not taken into account for the triple population here) and the double degenerate channel (circular CO WD binary merger due to GW emission with $m_\mathrm{IB} > M_\mathrm{Ch}$). Fig. \ref{figDTDcomp} shows that the triple channel DTD generally follows the same shape as the binary population synthesis DTD, with a rate proportional to $t^{-1}$ for times later than $t \approx 0.4 \, \mathrm{Gyr}$ (the best fitting slope for the triple case for $t > 0.4 \, \mathrm{Gyr}$ is $\approx -1.08$). In absolute terms, however, the triple-induced rates are much lower than the binary population synthesis rates: the former are typically factors of $\sim 10^2 - 10^3$ lower than the binary population synthesis rates. The total time-integrated rates are about $4 \cdot 10^{-4} \, M_\odot^{-1}$ and $2 \cdot 10^{-6} \, M_\odot^{-1}$ for the binary and triple cases respectively while the observed integrated rate from \citet{maozmanbrandt12} is $(1.30 \pm 0.15) \cdot 10^{-3} \, M_\odot^{-1}$. The triple-induced channels discussed here are therefore not able to resolve the discrepancy between the predicted and observed SNe Ia rates (see also \citealt{maozman12} and \citealt{wanghan12}). 

It is important to emphasize, however, that the present study does not give a complete picture of the triple-induced rate because we have not taken into account triple systems with inner binary orbits with $l_1 < 12 \, \mathrm{AU}$. Further study is therefore needed to examine the contributions from triple systems with $l_1 < 12 \, \mathrm{AU}$.

\section{Discussion}
\label{sect:discussion}
\subsection{Accelerating CO WD mergers through Kozai cycles}
\label{sect:discussion:acc}
In Sect. \ref{sect:introduction} we mentioned the scenario of \citet{thomp11} in which the tertiary induces high-amplitude Kozai cycles in close CO WD binary systems, thus reducing the GW merger time and producing a GW signature that is distinct from the signature from GW mergers in circular orbits. \citet{thomp11} considers triples with close inner binary CO WD systems ($10^{-2} < a_1 / \mathrm{AU} < 10$) and the outer orbit is assumed to be fairly tight ($10 \lesssim a_2/a_1 \lesssim 100$). For coeval triple systems it is also important to take into account the evolution prior to the formation of the binary CO WD system. Such evolution likely involves CE evolution in the inner binary system, which significantly shrinks the inner orbit and widens the outer orbit, thus substantially increasing the semi-major axis ratio $\beta \equiv a_2/a_1$. For the coeval triples considered in this work we find that the semi-major axis ratio $\beta$ is large enough ($\beta \sim 10^5$, cf. Fig. \ref{fig:betaCE}) to completely damp any subsequent Kozai cycles because of general relativistic orbit precession in all systems in which a close CO WD system is formed after a CE. We thus find that the scenario of \citet{thomp11} does not occur in coeval triples with $l_1 > 12 \, \mathrm{AU}$ (the systems with $l_1 < 12 \, \mathrm{AU}$ have not been modeled here). 

Furthermore we briefly mention the very recent work of \citet{katz12} who find that collisions of compact objects can occur in triple systems in which high eccentricities are reached through Kozai cycles. This scenario is very similar to the eccentric merger scenario found in this study (cf. Sect. \ref{sect:results:mergers:compobj}). \citet{katz12} use direct $N$-body methods to model the dynamical evolution of triple WD systems and find that the secular description of Kozai cycles breaks down in many of their systems. We note, however, that \citet{katz12} only consider equal-mass inner binaries, whereas we find that most eccentric WD collisions occur in systems with $q_1 < 1$ (Fig. \ref{figCOCOmtotq}, bottom panel) in which octupole-order terms dominate the dynamics. In addition, \citet{katz12} do not take into account stellar, binary and triple evolution prior to the formation of the double WD system. We find that these aspects are very important, however: most systems with high $a_1$, low $\beta \equiv a_2/a_1$ and high inclination, i.e. potential progenitors for the collision scenario, experience an early merger already during the MS prior to evolving to systems with compact objects (cf. Sect. \ref{sect:results:mergers:MS}). 

\subsection{Uncertainties in the triple evolution algorithm}
\label{sect:discussion:uncertainties}
In the triple algorithm developed in this work (Sect. \ref{sect:triplealg}) various simplifying assumptions are made, some of them not generally valid. Most importantly, we assumed a fast and isotropic wind mass loss for the outer binary orbit. If the inner orbit is tight and the inner orbital speeds are fast compared to the stellar wind speeds then the wind from either component in the inner binary system is likely affected by its inner binary companion (e.g. \citealt{valbor09}), thus invalidating our assumption. In our triple population this could be the case for some of the tighter inner binary systems containing AGB stars (with typically slow winds), in which case strong wind interaction is possible even in orbits with semi-major axes exceeding 12 AU (e.g. \citealt{moh07}). Because our results are strongly dependent on this assumption of fast and isotropic wind mass loss it would be worthwhile to implement a more sophisticated treatment of mass loss in our triple algorithm. A similar problem arises whenever a CE is invoked: in the present study it is assumed that the CE material from the inner binary is lost as if it were a fast and isotropic wind from the inner binary system. 

As Fig. \ref{fig:fracinitdep} shows our results are strongly dependent on the assumed triple distributions which are generally poorly constrained. Although the results from the two populations TSM1 and TSM2 are very similar (Table \ref{table:outcomes}), it must be noted that in both populations the initial inclination angle is sampled from the same distribution (i.e. a uniform distribution in $\cos[i_\mathrm{tot}]$). More observational constraints are needed on the initial triple parameters, in particular $i_\mathrm{tot}$. 

Furthermore, we assumed that the secular three-body dynamics, tidal friction and GW emission can be described in concert by linearly adding the relevant ODE terms (Eq. \ref{eq:tripleODEs}). It would be of value to investigate the validity of this assumption by means of more detailed simulations where a direct three-body code is coupled with hydrodynamics and/or stellar evolution codes, e.g. with the AMUSE framework \citep{SPZ12}. 

Another point of concern is whether our secular approach to hierarchical triple systems is still valid in the limit of the extreme eccentricities found in the population synthesis study (for eccentric CO WD mergers typically $1-e_1 \sim 10^{-7}$, cf. Sects. \ref{sect:extrev:TIEM} and \ref{sect:results:mergers:compobj}). To investigate this we have computed the gravitational dynamics for a system in our population synthesis sample after the formation of the triple CO WD system with a high-precision direct $N$-body code (Boekholt, in prep.) and we indeed find such high eccentricities. 

In addition, we note that uncertainties remain in the physics of tidal friction, in particular in the dynamical tide model where high eccentricities may induce complicated coupling of stellar oscillations with the tidal potential \citep{witsav00,zahn05}, thus possibly making the tidal strength quantities $k_{\mathrm{am},i}/T_i$ \citep{hut81,hur02} dependent on orbital eccentricity. We have investigated possible consequences of this scenario by means of an ad hoc approach, increasing the tidal strength quantities $k_{\mathrm{am},i}/T_i$ by a factor of $10^3$ separately for the cases of radiative and degenerate damping and rerunning the calculations. We find that this makes no significant difference.

Lastly, a possible concern is the fact that in our treatment of the secular dynamics of hierarchical triple systems not all post-Newtonian (PN) terms of the lowest order, i.e. the 1PN terms $\propto (v/c)^2$, are included in the Hamiltonian. As shown recently by \citet{naoz12}, a correct derivation starting from the three-body Hamiltonian accurate to 1PN order introduces an additional interaction term in the secularly averaged Hamiltonian. In an additional run of the calculations we have taken this interaction term into account and we find no significant difference.

\section{Conclusions}
\label{sect:conclusions}
We have performed a population synthesis study of triples with inner binary systems having initially $l_1 \equiv a_1 \left(1-e_1^2 \right) > 12 \, \mathrm{AU}$ and a primary mass range $1.0 - 6.5 \, M_\odot$. Our main conclusions are as follows. 

\noindent 1. In the considered triple sample the inner binary system does not interact in the absence of the tertiary. However, with the tertiary present the inner binary evolution is significantly affected in about $24\%$ of all simulated systems. As expected the process of Kozai cycles with tidal friction (KCTF) can be very important during the primary giant phases (RGB/AGB) whereas this was not the case during the main sequence. KCTF during the giant phases can drive strong orbital shrinkage and is often followed by common envelope evolution. As a consequence many different merger scenarios are possible.

\noindent 2. After a common envelope driven by KCTF and radius expansion further Kozai cycles are suppressed by general relativistic precession in the inner orbit. For close white dwarf binary systems formed in this scenario, this excludes the possibility of significantly reducing the gravitational wave merger time owing to high-amplitude Kozai cycles, as suggested by \citet{thomp11}. In the scenario of the present work the evolution prior to the white dwarf formation is strongly influenced by the tertiary, however. 

\noindent 3. In some systems with wide inner binaries KCTF is avoided during the giant phases, but a subsequent decrease in the semi-major axis ratio $a_2/a_1$ due to mass loss drives Kozai cycles with extreme eccentricity amplitudes. A subset of these systems is expected to experience an orbital collision between two carbon-oxygen white dwarfs in the inner binary system, which is a novel candidate SNe Ia progenitor scenario not found in any binary evolution path.

\noindent 4. We have identified three triple-induced carbon-oxygen white dwarf merger channels in our population synthesis study, among which the novel eccentric collision scenario, and have estimated the expected SNe Ia rates resulting from such mergers as function of time. We have compared these results to the binary population synthesis study of Claeys et al. (in prep.) and to the observations and find that the contribution from the triple-induced channels is small. We point out that further study is necessary in which triples with $l_1 < 12 \, \mathrm{AU}$ are also taken into account. However, in this work we have determined a lower limit to the contribution from triples to the SNe Ia rate.

\section*{Acknowledgements}
AH would like to thank T. Boekholt for verifying the gravitational dynamical evolution of a triple system in our population synthesis sample with his high-precision direct $N$-body code and the referee for providing helpful comments. Also AH would like to thank RU Nijmegen for hospitality during visits. This work was supported by the Netherlands Research Council NWO (grand 
\#639.073.803 [VICI]).

\bibliographystyle{mn2e}
\bibliography{literature}

\appendix

\section{DTD normalization calculations}
\label{app:DTDcalc}
We describe our method to normalize the delay time distribution (DTD) of Sect. \ref{sect:implications:rates} to the part of the mass represented by the sampled triple systems. Note that for the purposes of a comparison of the triple-induced rates to those of the binary population synthesis study of Claeys et al. (in prep.) a restriction must be made to the TSM1 population which is based on similar (inner) binary distributions. In this work the triple populations are split into two parts with $1.0 < m_1 /M_\odot < 2.0$ (TSM1A) and $2.0 < m_1/M_\odot < 6.5$ (TSM1B) (cf. Sect. \ref{sect:popsyn:samplmeth}). Therefore we perform the calculations separately for these separate mass ranges; subsequently we add the contributions from both mass ranges. 

We decompose the galactic population of $N_\mathrm{tot}$ gravitationally bound stellar systems into $N_\mathrm{bin} = \alpha_\mathrm{bin} N_\mathrm{tot}$ binary star systems, $N_\mathrm{tr} = \alpha_\mathrm{tr} N_\mathrm{tot}$ triple star systems and $N_\mathrm{s} = (1-\alpha_\mathrm{bin} - \alpha_\mathrm{tr}) N_\mathrm{tot}$ single star systems (i.e. we neglect higher-order multiplicities). We disregard from the calculated triple systems in TSM1 the MS mergers and MS destabilizations because these are not expected to be part of the observed population. There thus remain $N_{\mathrm{calc},\mathrm{A}} = 915,395$ and $N_{\mathrm{calc},\mathrm{B}} = 923,660$ computed systems for the two mass ranges respectively. These remaining triple systems constitute a fraction $f_\mathrm{calc}$ of all $N_\mathrm{tr}$ triple systems such that $N_\mathrm{calc} = f_\mathrm{calc} N_\mathrm{tr}$. Assuming that the initial primary mass $m_1$ is uncorrelated with $a_1$ and $e_1$, $f_\mathrm{calc} = f_{\mathrm{calc},m_1} \times f_{\mathrm{calc},e_1,e_2,a_1,a_2}$, where $f_{\mathrm{calc},m_1}$ is the fraction of calculated systems to all systems with respect to the mass distribution and $f_{\mathrm{calc},e_1,e_2,a_1,a_2}$ is this fraction with respect to the eccentricity and semi-major axes distributions of the inner and outer orbits. For a \citet{kroupa93} initial mass distribution that is used in TSM1,
\begin{align}
\frac{\mathrm{d}N}{\mathrm{d}m} \propto \left \{ 
\begin{array}{ll}
m^{\alpha_1}, & m_{\mathrm{Kr},1} < m < m_{\mathrm{Kr},2}; \\
m^{\alpha_2}, & m_{\mathrm{Kr},2} < m < m_{\mathrm{Kr},3}; \\
m^{\alpha_3}, & m_{\mathrm{Kr},3} < m < m_{\mathrm{Kr},4}, \\
\end{array} \right.
\label{eq:kroupaimf}
\end{align}
where $\alpha_j = \{-1.3,-2.2,-2.7\}$ and $m_{\mathrm{Kr},j} /M_\odot = \{0.1,0.5,1,80\}$, the fraction of systems with $m_{1,l} < m_1 < m_{1,u}$ is given by: \\
\begin{align}
f_{\mathrm{calc},m_1} = \frac{1}{\alpha_3+1} \, C_\mathrm{Kr} C_{\mathrm{Kr},3} \left (m_{1,u}^{\alpha_3+1} - m_{1,l}^{\alpha_3+1} \right ),
\label{eq:fm1sym}
\end{align}
where 
\begin{align}
C_\mathrm{Kr} = \left \{ \sum_{j=1}^3 \frac{C_{\mathrm{Kr},j}}{\alpha_j+1} \left (m_{\mathrm{Kr},j+1}^{\alpha_j+1} - m_{\mathrm{Kr},j}^{\alpha_j+1} \right ) \right \}^{-1}
\end{align}
and $C_{\mathrm{Kr},j} \equiv \{1, m_{\mathrm{Kr},2}^{\alpha_1-\alpha_2},m_{\mathrm{Kr},2}^{\alpha_1-\alpha_2} \, m_{\mathrm{Kr},3}^{\alpha_2-\alpha_3}\}$. Here $m_{1,l} = 1.0 \, M_\odot$  and $m_{1,u} = 2.0 \, M_\odot$ for TSM1A and $m_{1,l} = 2.0 \, M_\odot$  and $m_{1,u} = 6.5 \, M_\odot$ for TSM1B. Evaluating Eq. \ref{eq:fm1sym} numerically for the two mass ranges A and B we find $f_{\mathrm{calc},m_1,\mathrm{A}} \approx 6.34 \cdot 10^{-2}$ and $f_{\mathrm{calc},m_1,\mathrm{B}} \approx 2.44 \cdot 10^{-2}$. Note that this implies relative weights of $0.722$ and $0.278$ for the two mass ranges respectively; these weights are used for the calculations in Sect. \ref{sect:results}. 

The determination of $f_{\mathrm{calc},e_1,e_2,a_1,a_2}$ is complicated by the fact that $a_1$ and $a_2$ are correlated because the stability criterion is used of \citet{mard01}. This criterion has a non-trivial dependence on $e_2$ and $q_2 \equiv m_3/(m_1+m_2)$. In addition, the selection $l_1 \equiv a_1 \left(1-e_1^2\right ) > l_{1,l} \equiv 12 \, \mathrm{AU}$ is made. In an analytic treatment one would therefore have to integrate the TSM1 probability density function associated with $f_{\mathrm{calc},e_1,e_2,a_1,a_2}$, $\mathrm{d}^4 N / \left ( \mathrm{d} e_1 \mathrm{d} e_2 \mathrm{d} a_1 \mathrm{d} a_2 \right ) \propto e_1 e_2 / (a_1 a_2)$, with respect to $e_1$, $e_2$, $a_1$ and $a_2$, where $a_1 \beta_\mathrm{crit} < a_2 < a_u$, $a_l/\left(1-e_1^2\right) < a_1 < a_u$, $0<e_1< \left (1-l_{1,l}/a_u \right )^{1/2}$ and $0<e_2<1$. Here $a_l = 5 \, R_\odot$ and $a_u = 5 \cdot 10^6 \, R_\odot$ are the lower and upper limits respectively for the semi-major axis distribution \citep{kouw07}; the upper limit of $e_1$ is a consequence of the requirement that $a_u \left(1-e_1^2\right) > l_{1,l}$ or equivalently $e_1 < e_{1,u} \equiv \left (1-l_{1,l}/a_u \right )^{1/2}$. The required integration is complicated to do analytically.

Instead of this analytic method we determine the fraction $f_{\mathrm{calc},e_1,e_2,a_1,a_2}$ by the following method: we generate a large sample of systems according to the method appropriate for TSM1A and TSM1B without making the selection $l_1 > l_{1,l}$. From these systems we subsequently select the systems that satisfy $l_1  > l_{1,l}$. The number of the latter systems divided by the number of sampled systems represents an estimate of $f_{\mathrm{calc},e_1,e_2,a_1,a_2}$ if a sufficiently large number of systems is sampled. We perform this procedure with sizes varying between $10^4$ and $10^6$ and thus find $f_{\mathrm{calc},e_1,e_2,a_1,a_2} \approx 0.130$ for both TSM1A and TSM1B (i.e. this fraction is independent of primary mass). This implies calculated fractions of $f_{\mathrm{calc},\mathrm{A}} \approx 8.24 \cdot 10^{-3}$ and $f_{\mathrm{calc},\mathrm{B}} \approx 3.17 \cdot 10^{-3}$. 

The last step is to add the mass contributions from the single, binary and triple star systems. In general, for a Kroupa IMF the total mass $M$ in a population of size $N$ is given by:
\begin{align}
\nonumber M &= \int_{m_{\mathrm{Kr},1}}^{m_{\mathrm{Kr},4}} m \frac{\mathrm{d}N}{\mathrm{d} m} \, \mathrm{d} m = N \times C_\mathrm{Kr} \sum_{j=1}^3 \frac{C_{\mathrm{Kr},j}}{\alpha_j+2} \left (m_{\mathrm{Kr},j+1}^{\alpha_j+2} - m_{\mathrm{Kr},j}^{\alpha_j+2} \right ) \\
& \equiv N M_\mathrm{Kr},
\end{align}
where we have defined a ``Kroupa mass'' $M_\mathrm{Kr} \approx 0.5006 \, M_\odot$ that represents the average mass of a single star in the population. Hence the total mass of the single star population is given approximately by $M_\mathrm{s} \approx M_\mathrm{Kr} N_\mathrm{s}$. For the binary population we take a flat mass ratio distribution (consistent with Claeys et al., in prep.) such that the secondary is on average half as massive as the primary. Therefore the total mass contained in the binary population, $M_\mathrm{bin}$, is given approximately by $M_\mathrm{bin} \approx \left(1 + \frac{1}{2}\right) M_\mathrm{Kr}  N_\mathrm{bin} = \frac{3}{2} M_\mathrm{Kr} N_\mathrm{bin}$. Similarly, using that in TSM1 the distribution of the outer orbit mass ratio $q_2 = m_3/(m_1+m_2)$ is flat the total mass contained in the triple population is $M_\mathrm{tr} \approx \left(\frac{3}{2} + \frac{1}{2} \cdot \frac{3}{2} \right ) M_\mathrm{Kr}  N_\mathrm{tr} = \frac{9}{4} M_\mathrm{Kr} N_\mathrm{tr}$. Thus the total mass $M_\mathrm{tot} = M_\mathrm{s} + M_\mathrm{bin} + M_\mathrm{tr}$ represented by the number of triple systems we have calculated ($N_\mathrm{calc} = f_\mathrm{calc} N_\mathrm{tr}$) is given by: 
\begin{align}
M_\mathrm{tot} &\approx M_\mathrm{Kr} \left(N_\mathrm{s} + \frac{3}{2} N_\mathrm{bin} + \frac{9}{4} N_\mathrm{tr} \right ) = M_\mathrm{Kr} \frac{N_\mathrm{calc}}{f_\mathrm{calc}} \frac{1}{\alpha_\mathrm{tr}} \left (1 + \frac{1}{2} \alpha_\mathrm{bin} + \frac{5}{4} \alpha_\mathrm{tr} \right ).
\label{eq:DTDnorm:totmass}
\end{align}
With a binary fraction of $\alpha_\mathrm{bin} = 0.60$ and a triple fraction of $\alpha_\mathrm{tr} = 0.25$ we thus find that the masses represented by TSM1A and TSM1B are given by $M_{\mathrm{tot},\mathrm{A}} \approx 3.59 \cdot 10^8 \, M_\odot$ and $M_{\mathrm{tot},\mathrm{B}} \approx 9.41 \cdot 10^8 \, M_\odot$. We use these masses to normalize the triple-induced DTDs in Sect. \ref{sect:implications:rates}.

\end{document}